\documentclass[aps,physrev,preprint,groupedaddress]{revtex4-2}

\usepackage{graphicx}
\usepackage{dcolumn}
\usepackage{bm}
\usepackage{amssymb}
\usepackage{amsmath}
\usepackage{orcidlink}

\def\b{\begin{equation}}
 \def\e{\end{equation}}

\newcommand{\dif}{\mathrm{d}}
\newcommand{\bq}{\begin{eqnarray*}}
\newcommand{\eq}{\end{eqnarray*}}
\newcommand{\beq}{\begin{eqnarray}}
\newcommand{\enq}{\end{eqnarray}}

\begin{document}

\title{\textbf{Isocurvature-induced features in multi-field Higgs-$R^2$ inflation.} }

\author{Flavio Pineda\orcidlink{0000-0002-1265-5229}} 
\email{Contact author: fpineda@xanum.uam.mx}

\affiliation{{Departamento de Física, Universidad Autónoma Metropolitana Iztapalapa},
            {Av. San Rafael Atlixco No. 186, Colonia Vicentina}, 
            {CDMX},
            {09340}, 
            {Mexico},
            {Mexico}}

\author{Luis O. Pimentel\orcidlink{0000-0002-3614-7237}}
\email{lopr@xanum.uam.mx}

\affiliation{{Departamento de Física, Universidad Autónoma Metropolitana Iztapalapa},
            {Av. San Rafael Atlixco No. 186, Colonia Vicentina}, 
            {CDMX},
            {09340}, 
            {Mexico},
            {Mexico}}

\date{\today}

\begin{abstract}
We study primordial perturbations in Higgs--$R^2$ inflation in the presence of non-minimal kinetic mixing between the Higgs field and the scalaron. By numerically solving the multi-field background and linear perturbation equations, we identify distinct dynamical regimes controlled by the Higgs non-minimal coupling $\xi_h$. For $\xi_h \sim \mathcal{O}(0.1)$, transient turning of the inflationary trajectory leads to a transfer between adiabatic and isocurvature modes, generating localized features in the primordial curvature power spectrum. In contrast, in the weak-coupling regime $\xi_h \ll 1$, the curvature spectrum remains nearly featureless while isocurvature perturbations do not fully decay, resulting in a residual isocurvature component at the end of inflation. We compute the associated CMB angular power spectra and discuss the observational implications of these regimes. Our results highlight the role of multi-field dynamics in shaping primordial perturbations and provide constraints on viable realizations of Higgs--$R^2$ inflation.
\end{abstract}

\maketitle

\section{Introduction}

Cosmic inflation provides a compelling framework for describing the early Universe, addressing the horizon and flatness problems while generating the primordial perturbations observed in the Cosmic Microwave Background (CMB) \cite{guth1981,Linde1982,Liddle2003}. In its simplest realization, inflation is driven by a single scalar field and predicts a nearly adiabatic and scale-invariant spectrum of curvature perturbations \cite{Stewart1993}, a picture strongly supported by current observations \cite{planck2018x,ACT_2025}. Nevertheless, realistic ultraviolet completions of inflation typically involve multiple scalar degrees of freedom, as is the case in supersymmetry, Grand Unified Theories, and string-theoretic constructions. As a result, multi-field inflationary scenarios arise naturally and generically predict the presence of isocurvature perturbations, non-trivial field-space geometries, and turning inflationary trajectories \cite{Wands2002}.

A distinctive feature of multi-field inflation is the coupled evolution of adiabatic and isocurvature modes. Depending on the background dynamics, isocurvature perturbations may decay rapidly, transfer power to the curvature mode, or persist until the end of inflation, potentially leaving observable imprints in primordial correlators \cite{Bartolo2001,Huston2012,Dimarco2003,MUKHANOV1998,Lalak2007,STAROBINSKY2001,Langlois1999,LANGLOIS2003}. 

These effects can lead to scale-dependent features in the scalar power spectrum, correlated adiabatic--isocurvature signals, and enhanced non-Gaussianity \cite{Saenz2020,Chen_2010,Battefeld_2007,Bernardeau_2002,Iarygina_2024,Kaiser2013a,Elliston_2012}. Well-studied realizations include models with non-minimal couplings \cite{Schutz2014,Kaiser2014,Kaiser2010}, kinetic mixing between scalar fields \cite{Braglia2020,Dimarco2003,DiMarco2005}, and scenarios involving heavy fields with transient excitations \cite{Achucarro2011,Sebastian2012,Gao2012}. In this sense, isocurvature perturbations provide a powerful diagnostic of the underlying multi-field dynamics of inflation. Within this broad class of models, Higgs--$R^2$ inflation stands out as a theoretically well-motivated and predictive framework \cite{EMA2017,He2018,Wang2017,GUNDHI2020}. By combining the Standard Model Higgs field with a curvature-squared term, the model naturally introduces an additional scalar degree of freedom (the scalaron) and raises the unitarity cutoff to the Planck scale, thereby restoring perturbative control \cite{BEZRUKOV2008,SALVIO2015,EMA2017,GORBUNOV2019}. Although Higgs--$R^2$ inflation is often analyzed in an effective single-field limit \cite{Wang2017,He2018,EMA2017,GUNDHI2020}, its fundamental description is intrinsically multi-field, characterized by a curved field-space metric and kinetic mixing between the Higgs and the scalaron. This structure makes the model an ideal laboratory to investigate the generation, transfer, and persistence of isocurvature perturbations in a realistic inflationary setting. Higgs--$R^2$ inflation has been explored in a wide range of contexts, including preheating \cite{BEZRUKOV2019657,He2021,He2019,He2021tachionic}, gravitational particle production and primordial magnetogenesis \cite{Pineda2025,Durrer2022}, cosmological collider signatures \cite{Ema2024}, gravitational waves \cite{Kim2025}, primordial black-hole formation \cite{Cheong2021,Cheong_2022, Pi_2018, Wang_2024}, and small-scale CMB phenomenology \cite{Tanmoy2025}. Despite this extensive literature, a systematic analysis of the simultaneous evolution of adiabatic and isocurvature perturbations beyond the single-field attractor regime remains relatively unexplored, particularly in regions of parameter space where no strong hierarchy between the fields exists \cite{Wang2017}.

In this work, we revisit the multi-field dynamics of Higgs--$R^2$ inflation with a focus on the evolution and observational consequences of isocurvature perturbations. We concentrate on the regime where the non-minimal Higgs coupling is small, $\xi_h \sim \mathcal{O}(0.1)$, such that the inflationary trajectory can deviate from the single-field valley and undergo transient turns as it evolves from the ridge at $h=0$ toward the minima \cite{GUNDHI2020}. These turns generate a non-vanishing bending rate, $\eta_\perp$, which couples adiabatic and isocurvature modes and can induce localized features in the primordial curvature spectrum. We identify two qualitatively distinct regimes: one in which isocurvature perturbations persist until the end of inflation ($\xi_h \ll 1$), and another in which transient turning efficiently transfers power to the curvature mode while suppressing isocurvature by the end of inflation ($\xi_h \gtrsim \mathcal{O}(0.1)$).
We compute the resulting primordial power spectra and illustrate how these different dynamical regimes lead to characteristic imprints on the CMB angular power spectra. Our goal is not to provide a best-fit model to current data, but rather to characterize the physical origin and scale dependence of isocurvature-induced effects in a well-motivated multi-field inflationary scenario. These results highlight that suppressing features in the curvature spectrum does not necessarily guarantee the elimination of isocurvature perturbations, placing non-trivial constraints on viable realizations of Higgs--$R^2$ inflation and motivating future observational probes sensitive to primordial isocurvature.

This paper is organized as follows. In Section~\ref{sec2} we review the Higgs--$R^2$ model in the Einstein frame and discuss the background evolution. Section~\ref{sec3} presents the formalism for linear perturbations in multi-field inflation. In Section~\ref{sec4} we present our numerical results for the primordial spectra, while Section~\ref{sec5} illustrates their impact on the CMB angular power spectra. Lastly, in Section~\ref{sec6} we show the preliminary results on primordial non-gaussianity in this model. We summarize our conclusions in Section~\ref{sec7}. Throughout this work we assume a spatially flat FLRW background,

\begin{equation}
\dif s^2 = -\dif t^2 + a^2(t)\,(\dif x^2 + \dif y^2 + \dif z^2),
\label{FLRW metric}
\end{equation}

and use natural units with $c = \hbar = G = M_p = 1$.

\section{Multi-field Higgs-$R^2$ inflation}
\label{sec2}

Our starting point is the Higgs-$R^2$ model \cite{EMA2017, GORBUNOV2019, GUNDHI2020} in the Jordan frame in its most general form, i.e., the theory of Higgs inflation \cite{BEZRUKOV2008} augmented by an $R^2$ term, whose action is given by

 \beq
\nonumber    S_J[g_{\mu\nu}\,,\mathcal{H}] = \int \dif^4 x \sqrt{-g_J}\left[ \dfrac{1}{2}\left(1 + 2\xi_h|\mathcal{H}|^2 \right) R_J + \xi_s R_J^2 - g_J^{\mu\nu}(D_\mu \mathcal{H})^\dagger (D_\nu \mathcal{H}) -\lambda|\mathcal{H}|^4 \right]\,,\\
 \enq

where the variables defined in the Jordan frame are denoted by the subscript $J$. The field $\mathcal{H}$ represents the Higgs doublet with a non-minimal coupling $\xi_h$, which is necessary for $\mathcal{H}$ to drive inflation, while $\lambda$ is the quartic coupling constant and $D_\mu\mathcal{H}$ is the covariant derivative of the SM Higgs sector, which includes interactions with the gauge fields:

\beq
D_\mu\mathcal{H} = \left(\partial_\mu - \dfrac{ig}{2}\tau_a\,W^a{}_\mu  + \dfrac{ig'}{2}B_\mu\right)\mathcal{H}\,.
\enq

The fields $W^a_\mu$ and $B_\mu$ represent the gauge fields of the $SU(2)$ and $U(1)$ groups, with their respective coupling constants $g$ and $g'$, while $\tau_a$ are the $SU(2)$ generators. It is customary to define the Higgs doublet $\mathcal{H}$ in the unitary gauge such that $\mathcal{H}$ is parameterized by a single field $h$:

\beq
\mathcal{H} = \dfrac{1}{\sqrt{2}}\begin{pmatrix}
    0\\
    h + v_\mathrm{ew}
\end{pmatrix}\,,
\enq

where $v_\mathrm{ew} \approx 246\,\mathrm{GeV}$ is the electroweak energy scale. In this gauge, and considering that $h \gg v_\mathrm{ew}$, the action of the model can be rewritten as \cite{EMA2017}

\beq
    S_J[g_{\mu\nu}\,,h] = \int \dif^4\sqrt{-g_J}\left[ \dfrac{1}{2}\left(1 +  \xi_h h^2\right)R_J + \xi_s R_J^2 - \dfrac{1}{2}g_J^{\mu\nu}(\partial_\mu h)(\partial_\nu h) - \dfrac{\lambda}{4}h^4  \right]\,,
    \label{Higgs-R2 action}
\enq

where we have omitted interactions with the gauge fields. It is well established that this model is a natural extension of Higgs inflation that solves the large coupling constant problem \cite{GORBUNOV2019}. The Higgs model requires a coupling constant $\xi_h \sim \mathcal{O}(10^4)$ to render the model compatible with CMB observables; however, it suffers from the strong coupling problem \cite{Barbon2009, Burgess2009, BEZRUKOV2008}. This problem arises because the Higgs field becomes strongly coupled at the electroweak scale, causing a loss of perturbative unitarity well below the usual Planck scale \cite{GORBUNOV2019}. The cutoff scale is $\Lambda_\text{UV} = M_p/\xi_h$, which casts doubt on the model's predictions. For $\xi_h \sim \mathcal{O}(10^4)$, the energy scale is of the order of $\Lambda_\text{UV} \sim 10^{14}$, which is lower than the energy scale during inflation. However, the inclusion of the $R^2$ term naturally raises the model's cutoff to the Planck scale, thereby restoring its perturbative nature \cite{EMA2017,GORBUNOV2019}. The model \eqref{Higgs-R2 action} features two degrees of freedom: the field $h$ and a scalar degree of freedom arising from $R^2_J$, known as the scalaron. Both fields are dynamic and contribute to the expansion history of the universe. This dynamics is more easily understood by rescaling the spacetime metric $g_{\mu\nu}$ via a conformal transformation, $g_{\mu\nu}\to \Omega^2(x)g_{\mu\nu}$, so that the action \eqref{Higgs-R2 action} adopts the standard Einstein-Hilbert form plus an action for two scalar fields $\phi$ and $h$ \cite{Kaiser2010a}. 

The conformal transformation that allow us to write \eqref{Higgs-R2 action} in the Einstein frame is $\Omega^2(x) =e^{-\alpha \phi}$, with $\alpha = \sqrt{2/3}$ and $\phi$ being the scalaron field \cite{EMA2017}. The resulting action is

\beq
S_J \to S_E = \int \dif^4 x\sqrt{-g}\left[\dfrac{1}{2}R - \dfrac{1}{2}g^{\mu\nu}(\partial_\mu \phi)(\partial_\nu \phi) - \dfrac{1}{2}e^{-\alpha\phi}g^{\mu\nu}(\partial_\mu h)(\partial_\nu h) - V(\phi\,,h)\right ]\,,
    \label{accion marco einstein}
\enq

where the scalar potential $V(\phi\,,h)$ depends on both fields in the Einstein frame and is defined as

\beq
V(\phi\,,h) = e^{-2\alpha\phi}\left[\dfrac{1}{16\xi_s}\left(e^{\alpha\phi} - 1 - \xi_h\,h^2 \right)^2 + \dfrac{\lambda}{4}\,h^4\right]\,.
    \label{potential Einstein}
\enq

The non-canonical kinetic term of $h$ defines a non-trivial field-space metric $G_{IJ}$ with non-vanishing components

\beq
    G_{\phi\phi} = 1\,,\quad G_{hh} = e^{-\alpha\phi}\,.
    \label{field space metric}
\enq

This type of model, featuring a non-canonical kinetic term and two scalar fields, is well known and has been widely studied in the literature \cite{Lalak2007, STAROBINSKY2001, Dimarco2003, DiMarco2005, Tolley_2010, Cremonini2011}, exploring various simple forms of the inflationary potential. By this, we refer to inflationary potentials that can typically be factorized as a sum of single-field potentials $V(\phi\,,\chi) = U(\phi) + W(\chi)$, or as a product $V(\phi\,,\chi) = U(\phi)W(\chi)$. The potential \eqref{potential Einstein} is more involved in this sense, as it is not factorizable either as a sum or a product of individual potentials depending solely on $\phi$ or $h$.
Furthermore, this model presents three parameters: $\lambda$, $\xi_h$, and $\xi_s$. According to \cite{EMA2017}, there are three cases that yield inflation at the energy scale $\phi \gg M_p$:

\begin{enumerate}
    \item $\xi_h >0\,,\xi_s >0\,, \lambda >0$
    \item $\xi_h <0\,, \xi_s > 0\,, \lambda >0$
    \item $\xi_h <0\,, \xi_s > 0\,, \lambda < 0$
\end{enumerate}

In all three cases, the inflationary dynamics effectively reduces to a single-field scenario for $\xi_h \gg 1$, as the isocurvature mode becomes heavy enough to be integrated out. This reduction in inflationary dynamics can be understood via the valley-approach, which we briefly demonstrate in Appendix \ref{appendix1}.
The first case corresponds to a mixture between the Higgs field $h$ and the scalaron $\phi$; the inflaton potential is given by \eqref{effective potential}. 
In the second case, inflation is driven solely by the scalaron $\phi$, while the third corresponds to the metastable electroweak vacuum. On the other hand, if $\xi_s \gg \xi_h^2/\lambda$, inflation is of the $R^2$-type, whereas in the opposite case, inflation is predominantly driven by the Higgs field $h$ \cite{EMA2017}. However, if no hierarchy exists among the parameters, it is ensured that both $\phi$ and $h$ contribute to the inflationary dynamics and the generation of perturbations for $\xi_h \gg 1$. This case is known as the standard parameter space of the Higgs-$R^2$ model, where multi-field effects are known to be absent \cite{ Wang2017,He2018, GUNDHI2020}.

\begin{figure}[t]
    \centering
    \includegraphics[width=0.48\linewidth]{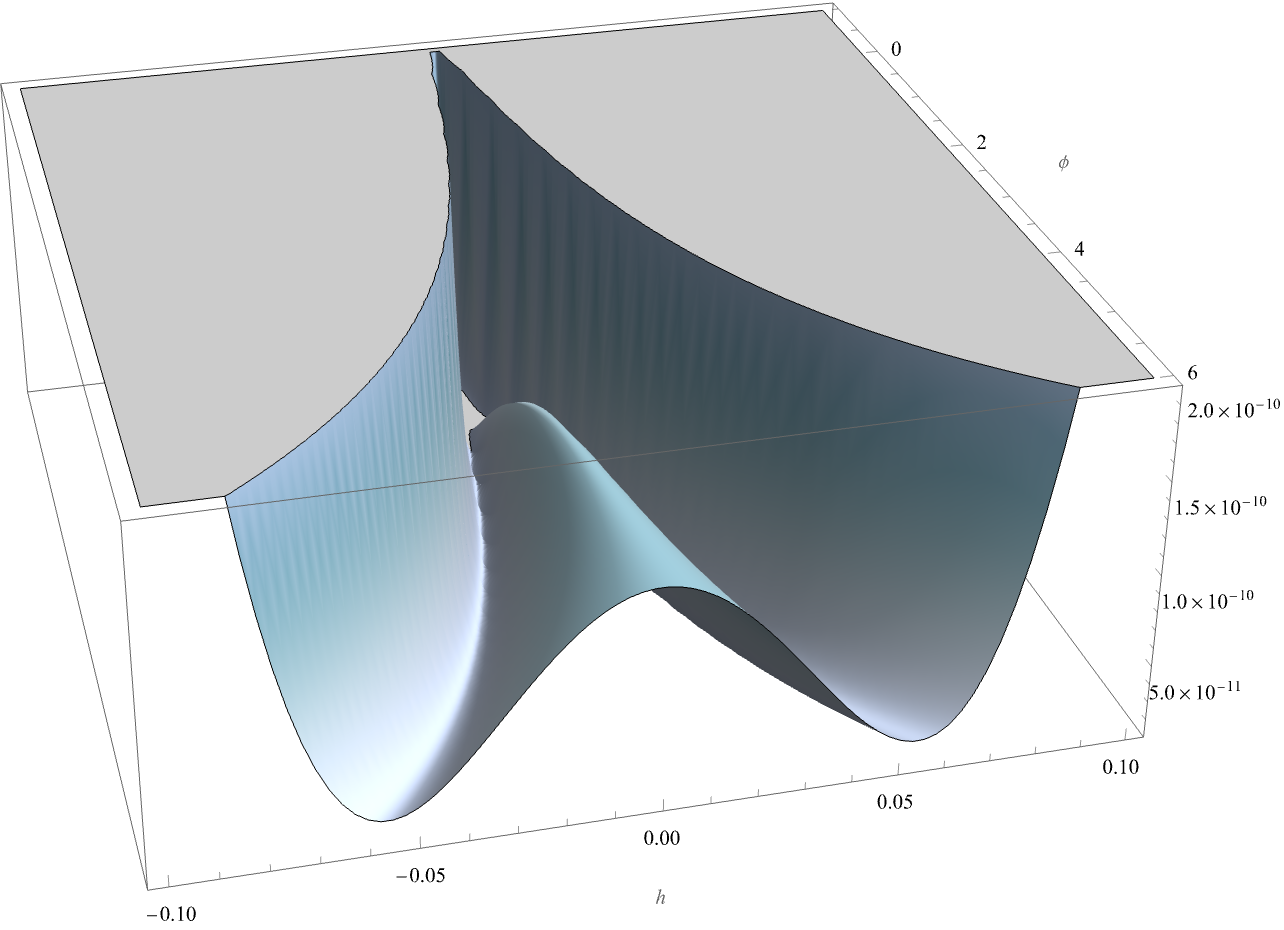}\quad
     \includegraphics[width=0.48\linewidth]{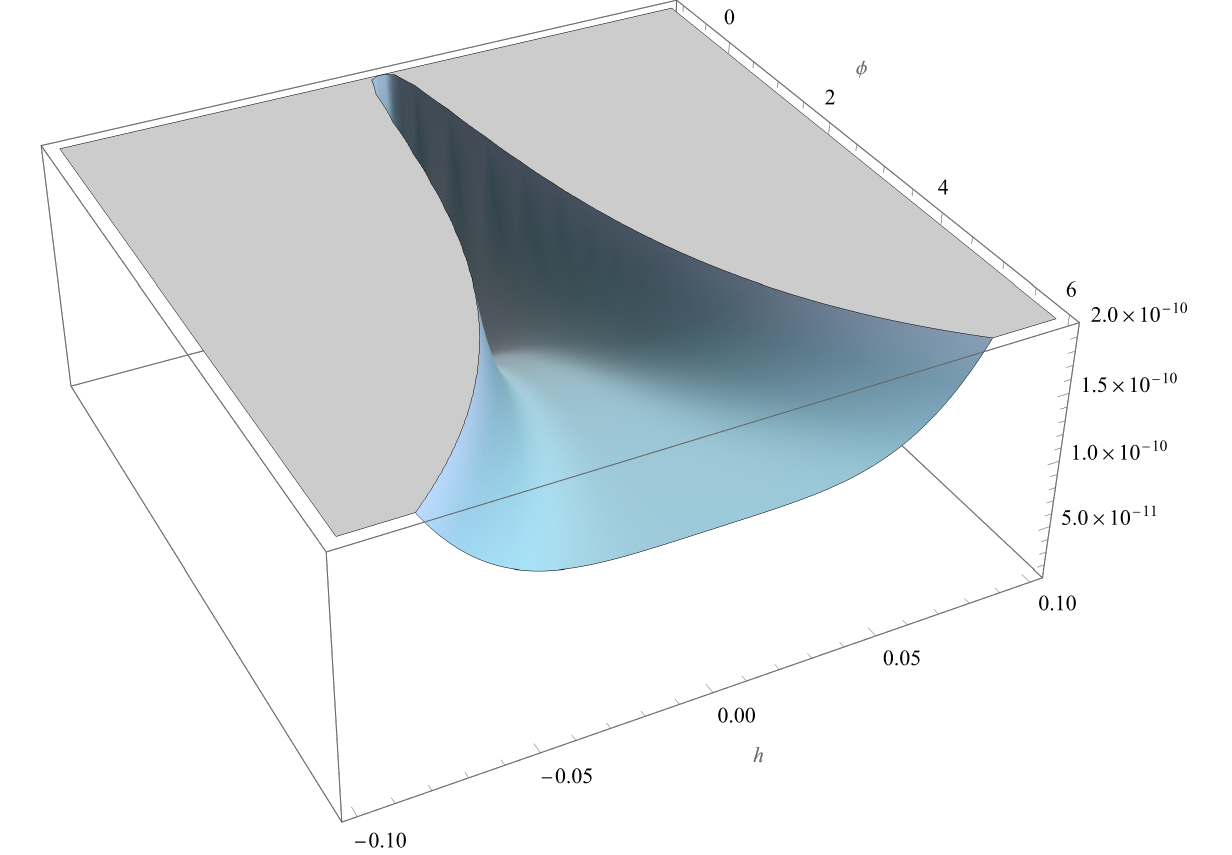}
    \caption{Scalar potential $V(\phi\,,h)/M_p^4$ in the Einstein frame. Left panel: Potential \eqref{potential Einstein} using the standard parameter space $\lambda = 0.13$, $\xi_h = 4000$ and $\xi_s \simeq 10^8 $. The potential exhibits a two-valley structure and a ridge at $h=0$. Right panel: The same potential \eqref{potential Einstein} but for a coupling $\xi_h = 0.1$. We observe that the valley and ridge structure disappears in the limit $\xi_h \ll 1$.}
    \label{fig1}
\end{figure}

In this work, we explore the case $\xi_h >0\,,\xi_s >0\,, \lambda >0$ with a non-minimal copupling $\xi_h $ satisfying $\xi_h \sim  \mathcal{O}(0.1)$ and $\xi_h \ll 1$,  such that isocurvature modes and their primordial features are not negligible during inflation. This condition ensures that the isocurvature mode is not too heavy to be excited during inflation, making its evolution relevant for primordial perturbations. The values of the parameters, beyond playing a role in the generation of isocurvature modes, impact the shape of the potential \eqref{potential Einstein}. In Fig. \ref{fig1}, we display the structure of the potential for two different parameter spaces. For a more detailed analysis of the potential shape and its consequences on field-space trajectories, we recommend Ref. \cite{GUNDHI2020}.

\subsection{Background dynamics}

To characterize the system's evolution in multi-field inflation models, it is customary to employ a covariant field-space formalism \cite{Achucarro2011, Gong2011, Gong2017, KARAMITSOS2018}. The equations of motion for the fields $\phi, h$ are derived directly from \eqref{accion marco einstein}. Using the covariant field-space approach and the metric \eqref{FLRW metric}, these are written as

\beq
    D_t\dot{\phi}^I + 3H\dot{\phi}^I + G^{IJ}V_J = 0\,,
    \label{EoM covariant}
\enq

where $D_t$ is the covariant directional derivative along $t$, defined acting on an arbitrary field-space vector $A^I$ as $D_t A^I = \dot{\phi}^J \nabla_J A^I = \dot{A}^I + \Gamma^I{}_{JK}\,\dot{\phi}^J\,A^K$, while the Christoffel symbols are evaluated in terms of $G_{IJ}$ and its derivatives via the expression $\Gamma^I{}_{JK} = G^{IM}(\partial_J G_{KM} + \partial_K G_{JM} -\partial_M G_{JK} )/2$, with $\partial_J G_{KM} = \partial G_{KM}/\partial \phi^J$. For the metric \eqref{field space metric}, the non-vanishing components of $\Gamma^I{}_{JK}$ are

\beq
\Gamma^\phi{}_{h h} = \dfrac{\alpha}{2}e^{-\alpha\phi}\,,\quad \Gamma^h{}_{\phi h} = \Gamma^h{}_{h \phi} = -\dfrac{\alpha}{2}\,.
\label{Christoffel}
\enq

In a spacetime given by the FLRW metric \eqref{FLRW metric}, and assuming that the scalar fields are homogeneous, $\phi = \phi(t)$ and $h = h(t)$, the equations of motion \eqref{EoM covariant} satisfied by the fields $(\phi, h)$ are given by

\begin{align}
 \ddot{\phi} + 3H\dot{\phi} + V_\phi = -\dfrac{\alpha}{2}e^{-\alpha\phi}\dot{h}\,,\quad
    \ddot{h} + (3H -\alpha\dot{\phi})\dot{h} + e^{\alpha\phi}\,V_h = 0\,,
    \label{EoM system}
\end{align}

where, as usual, an overdot denotes $\partial_t$, the subscripts on the potential indicate derivatives, and $H(t)$ is the Hubble parameter given by $H= \dot{a}/a$, with $a(t)$ being the scale factor. The evolution and dynamics of $H$ are governed by the Friedmann equations:

\begin{align}
\dot{H} = - \dfrac{1}{2} \dot{\sigma}^2 \,,\quad
 3 H^2 = \dfrac{1}{2}\dot{\sigma}^2 + V(\phi\,,h)\,,
 \label{Friedmann ecs}
\end{align}

where $\dot{\sigma}^2 = G_{IJ}\dot{\phi}^I\dot{\phi}^J =  \dot{\phi}^2 + e^{-\alpha\phi}\dot{h}^2$. The numerical solution of Eqs. \eqref{EoM system} and \eqref{Friedmann ecs}, with an appropriate choice of initial conditions, provides the background trajectory. To discuss the features of said trajectory and the field-space geometry, it is useful to define the tangent vector $T^I$ and the normal vector $N^I$ to the trajectory as $T^I = \dot{\phi}^I/\dot{\sigma}$ and $N^I =\sqrt{\text{det} G}\,\epsilon_{IJ} T^J$, where $\epsilon_{IJ}$ is the fully antisymmetric Levi-Civita symbol, and $\det G$ is the determinant of the metric \eqref{field space metric}. Explicitly, for the model \eqref{Higgs-R2 action}, these vectors are given by

\beq
    T^I = \dfrac{(\dot{\phi}\,,\dot{h})}{\dot{\sigma}}\,,\quad N^I = \dfrac{e^{\alpha\phi/2}}{\dot{\sigma}}\,( -e^{-\alpha\phi}\dot{h}\,,\dot{\phi})\,.
    \label{orthogonal vectors}
\enq

The vector $T^I$ defines the direction parallel to the trajectory (adiabatic) of any field-space vector $A^I$, while $N^I$ defines the first normal component (isocurvature) of $A^I$ \cite{Achucarro2011}. In this way, these vectors allow us to decompose any vector $A^I$ into its adiabatic component $A_\sigma$ and isocurvature component $A_s$ as $A^I = A_\sigma T^I + A_s N^I$. Furthermore, the vector $T^I$ offers an alternative way to write the equation of motion \eqref{EoM covariant} as two independent equations: $\ddot{\sigma} + 3H\dot{\sigma} + V_\sigma = 0$ and $D_t T^I = - H\eta_\perp N^I$, where we have defined $V_\sigma = T^I V_I$ and $V_N = N^I V_I$ as the projections of $\partial_I V = V_I$ along the adiabatic and isocurvature directions, respectively. The parameter $\eta_\perp$ is known as the turning rate and indicates the rate of change of $T^I$ along the $N^I$ direction.  If $\eta_\perp = 0$, the vectors \eqref{orthogonal vectors} remain covariantly constant along the classical trajectory, whereas if $\eta_\perp \neq 0$, the vectors $T^I$ and $N^I$ can rotate to the right or left, depending on the sign of $\eta_\perp$. It is important to note that if the trajectory $\phi^I(t)$ is a geodesic in field space ($D_t \dot{\phi}^I = 0$), the turning rate vanishes. 

Moreover, this parameter is responsible for sourcing the mixing between the adiabatic and isocurvature modes of the primordial perturbations and parameterizes the interaction between them; thus, for geodesic motion in field space, both perturbations evolve independently (see Section \ref{sec3}). Characterizing the dynamics of this type of model involves a slow-roll analysis analogous to that of single-field inflation models. We can define the slow-roll parameters as

\beq
    \epsilon = -\dfrac{\dot{H}}{H^2}\,,\quad \eta^I = - \dfrac{D_t\,\dot{\phi}^I}{H\dot{\sigma}}\,,
    \label{SR parameters}
\enq

where $\epsilon$ is the usual slow-roll parameter, while $\eta^I$ is a vector in field space containing information about the second slow-roll parameter. We decompose $\eta^I$ into its adiabatic and isocurvature components $\eta^I = \eta_{||} T^I + \eta_\perp N^I$, where $\eta_{||} = \eta^I T_I$ and $\eta_\perp = \eta^I N_I$ are given by

\beq
    \eta_{||} = -\dfrac{\ddot{\sigma}}{H\dot{\sigma}}\,, \quad \eta_\perp = \dfrac{V_N}{H\dot{\sigma}}\,.
    \label{second SR}
\enq

It is clear that the component $\eta_{||}$ is the counterpart of the second slow-roll parameter in single-field models, while $\eta_\perp$ is the turning rate parameter appearing in the equation $D_t T^I = - H\eta_\perp N^I$.
The slow-roll conditions in these models are given by $\epsilon\ll 1$ and $|\eta_{||}|\ll 1$, which is equivalent to $\ddot{\sigma} \ll |H\dot{\sigma}|$. These conditions do not imply a small turning rate $\eta_\perp \ll 1$; therefore, $\eta_\perp \gtrsim 1$ is compatible with the slow-roll approximation. Using the explicit form of $N^I$, we can construct the projection of the gradient of the potential \eqref{potential Einstein} along the isocurvature direction:

\begin{align}
    V_N = N^I V_I  = \dfrac{e^{\alpha\phi/2}}{\dot{\sigma}}(V_h \dot{\phi} - e^{-\alpha\phi}\dot{h} V_\phi)\,.
    \label{V_N}
\end{align}

\begin{figure}[t]
    \centering
    \includegraphics[width=0.48\textwidth]{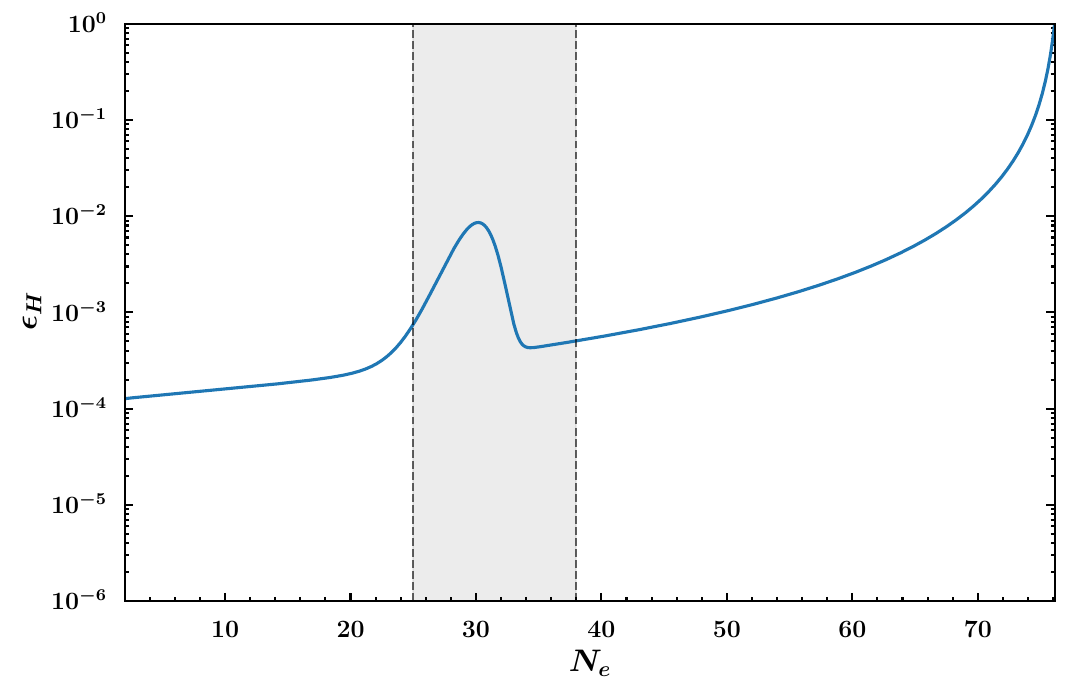}
    \quad
    \includegraphics[width=0.48\textwidth]{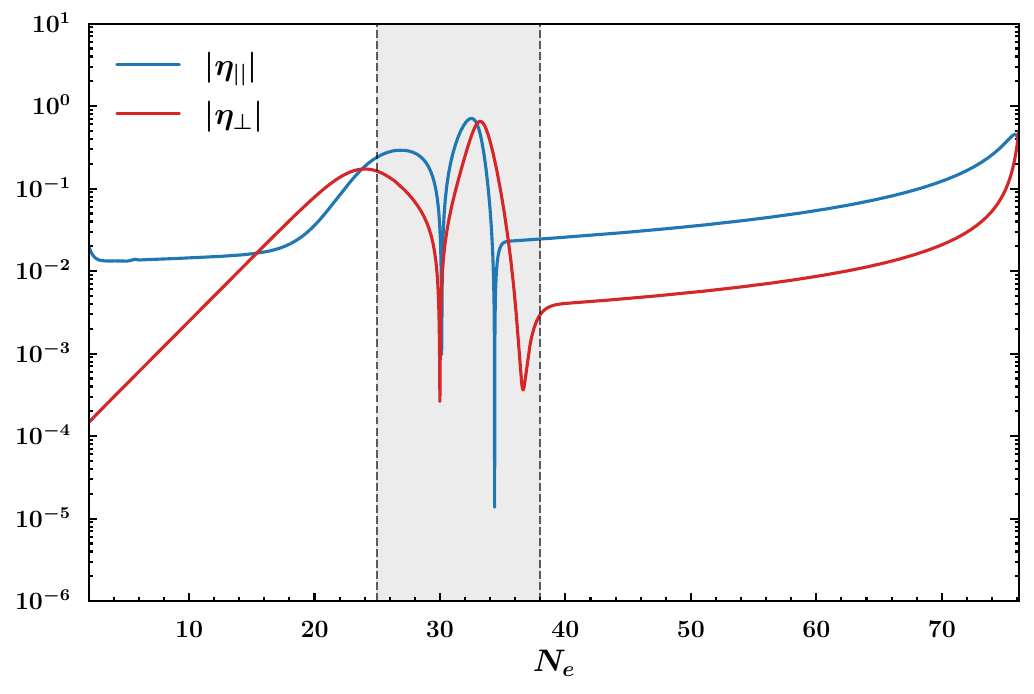}\\[0.3cm]
    \includegraphics[width=0.48\textwidth]{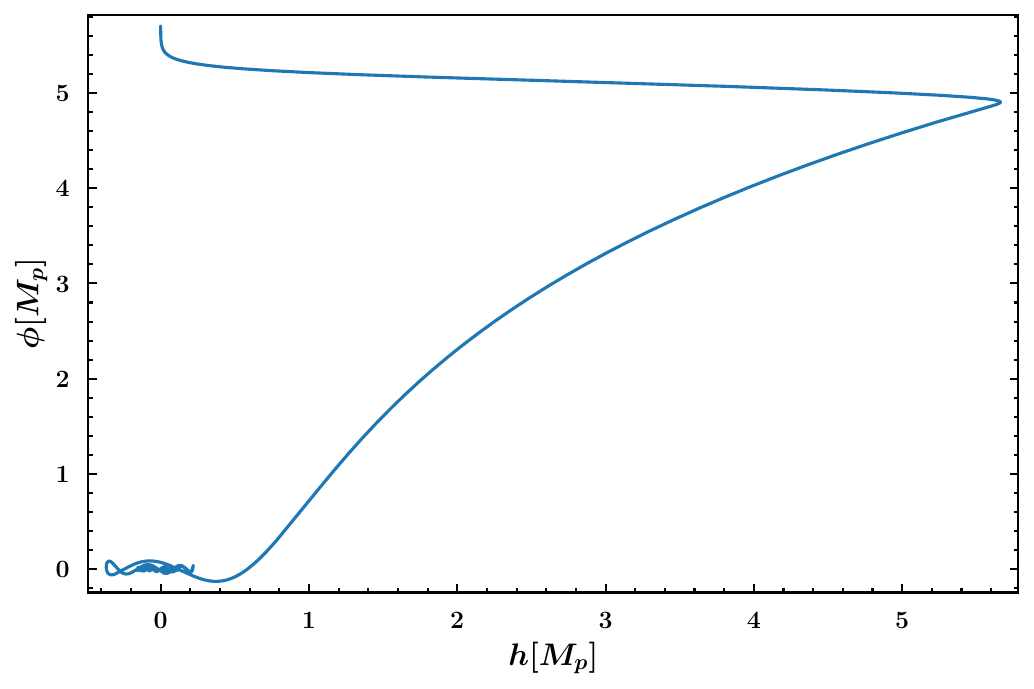}
    \quad
    \includegraphics[width=0.48\textwidth]{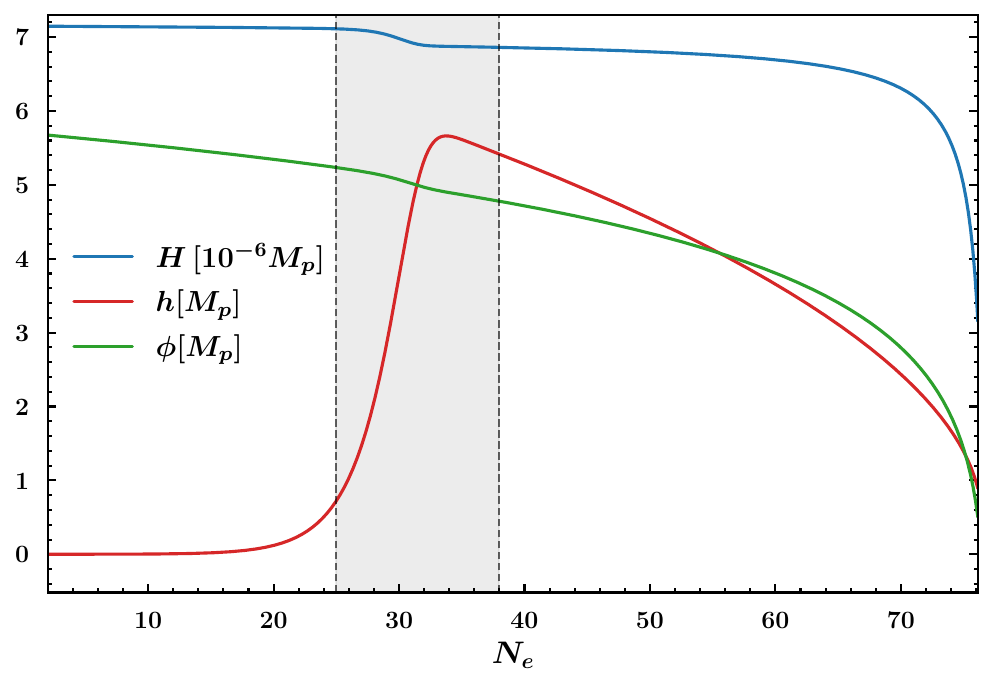}

    \caption{
    Evolution of the slow-roll parameter $\epsilon_H$, the turning rate parameter $\eta_\perp$, and the parameter $\eta_{||}$,
    along with the field-space trajectory and background variables
    ($H$, $\phi$, $h$) as a function of the number of e-folds $N_e$.
    We consider the parameter set $\xi_h = 0.1$, $\lambda = 10^{-10}$, and $\xi_s = 4\times10^8$,
    with initial conditions $\phi_0 = 5.7$ and $h_0(\phi_0) = 10^{-4}$ in Planck units.
    It is observed that, although inflation begins in a slow-roll regime ($\epsilon_H \ll 1$), the initial conditions produce transient peaks associated with turns in the field-space trajectory.}
    \label{Fig2}
\end{figure}

This allows us to write the model's turning rate as

\beq
\eta_\perp = \dfrac{e^{\alpha\phi/2}}{H\dot{\sigma}^2}\left( V_h \dot{\phi} - e^{-\alpha\phi}\,\dot{h}V_\phi \right)\,.
\label{turning rate}
\enq

Within the valley-approach, the turning rate and the orthonormal vectors to the field-space trajectory are given in \eqref{valley approach}. In the standard parameter space and within the valley-approach, the turning rate is negligible,  but for the parameter space satisfying $\xi_h \sim \lambda\xi_s$, the turning rate is of the order $\eta_\perp \sim \mathcal{O}(0.1)$. Nevertheless, for a proper understanding of the background evolution, it is convenient to perform a numerical analysis of both the background variables and the perturbations. To this end, it is convenient to use the number of e-folds $N_e$ as the time parameter, given the relation $\dif N_e = H \dif t$. In this case, equations \eqref{EoM system} and \eqref{Friedmann ecs} take the form

\begin{align}
    \phi'' + (3-\epsilon)\,\phi' + \frac{V_\phi}{H^2} &= -\frac{\alpha}{2} e^{-\alpha\phi} h',\nonumber\\[0.3cm]
    h'' + (3-\epsilon-\alpha\phi')\,h' + e^{\alpha\phi}\frac{V_h}{H^2} &= 0,\nonumber\\[0.3cm]
    H^2 &= \frac{V}{3-\epsilon}\,.
    \label{efolds system}
\end{align}

where a prime denotes $\partial_{N_e}$, while the slow-roll parameter $\epsilon$ in terms of e-folds is written as

\beq
    \epsilon = - \dfrac{H'}{H} = \dfrac{1}{2}(\phi'^2 + e^{-\alpha\phi}h'^2)\,,
    \label{efolds SR}
\enq

and the turning rate \eqref{turning rate} as a function of $N_e$ is given by

\beq
    \eta_\perp = \dfrac{e^{\alpha\phi/2}}{\sigma_N^2}\left( V_h\phi' - e^{-\alpha\phi}\,V_\phi h' \right)\,,
    \label{turning rate efolds}
\enq

where $\sigma_N^2 = H^2(\phi'^2 + e^{-\alpha\phi}h'^2)$. Once the model parameters and suitable initial conditions have been specified, we can proceed to numerically integrate the system of equations \eqref{efolds system} to obtain the relevant background quantities. With these quantities, we can proceed to analyze the primordial perturbations of the model.
Following Ref. \cite{Fumagalli2021}, the analysis and evolution of linear primordial perturbations are determined by three functions: the Hubble parameter $H(N)$, the turning rate $\eta_\perp(N)$, and the mass of the isocurvature mode $m_\text{iso}^2(N)$. In Fig. \ref{Fig2}, we display the numerical solution for the background; specifically, we show the behavior of $\epsilon_H(N)$, $\phi(N)$, $h(N)$, $H(N)$, $\eta_\perp(N)$, and $\eta_{||}(N)$ given the initial condition $\phi_0 = 5.7$ in Planck units. The initial condition of the field $h$ determines the starting position on the potential. If $h_0 = h_0^\text{valley}$ is given by condition \eqref{valley}, inflation begins exactly within one of the potential valleys; if $h_0 \ll 1$, multi-field effects associated with the fall from the ridge at $h=0$ towards one of the valleys are guaranteed \cite{GUNDHI2020}. As an example, we show the background for the initial condition $h_0 = 10^{-4}$ in Planck units. We can observe some important features in the background behavior.
Around $N_e \simeq 25-35$, a turn in the field-space trajectory occurs, which is reflected in the parameters $\eta_\perp$ and $\epsilon_H$. The small increase in $\epsilon_H$ represents a temporary deviation from the slow-roll regime associated with this turn.
Although inflation is not interrupted ($\epsilon_H \ll 1$ at all times), this behavior can leave observable imprints on the primordial power spectrum of $\mathcal{R}_k$ (see Section~\ref{powerspectrum}).
This type of behavior is typical of potentials with local features ("bumps" or "dips") designed to induce ultra-slow-roll phases ($\epsilon_H \to 0$) and amplify the primordial spectrum to values of $\mathcal{P}_\mathcal{R}(k) \sim 10^{-2} - 10^{-1}$, which are necessary for the formation of primordial black holes (PBHs) \cite{Mishra_2020}. Such mechanisms can also occur in multi-field models with transient turns in field space \cite{Braglia2020, Cheong2021, Cheong_2022}; however, in our case, the change in trajectory is not sharp enough to amplify the spectrum up to the PBH production regime.

\section{Primordial perturbations}
\label{sec3}

When studying primordial perturbations at linear or quadratic order in multi-field inflation models, it is advisable to adopt a covariant approach with respect to gauge transformations in field space. This formalism was developed in detail in Ref. \cite{Gong2011} for an arbitrary field space. In this paper, we neglect vector perturbations and focus primarily on scalar primordial perturbations.
From a covariant perspective, the strategy is to employ the covariant perturbation $Q^I$ instead of the coordinate perturbation $\delta\phi^I(t\,,\mathbf{r}) = \phi^I(t\,,\mathbf{r}) - \bar{\phi}^I(t)$, where $\bar{\phi}^I(t)$ is the classical background trajectory satisfying \eqref{EoM covariant}. The values of $\bar{\phi}^I$ and $\phi^I$ at a fixed spacetime point $P$ can be connected via a unique geodesic in field space parameterized by $\varepsilon$, which relates the initial value of $\phi^I$ to the tangent vector at $P$. The initial conditions are set at $\varepsilon = 0$, such that

\beq
    \phi^I (\varepsilon = 0) = \bar{\phi}^I\,,\quad Q^I = \left.\dfrac{\dif \phi^I}{\dif \varepsilon}\right|_{\varepsilon = 0} \,.
    \label{IC field space}
\enq

At second order in perturbations, one finds the expansion relating $Q^I$ to the perturbation $\delta\phi^I$:

\beq
    \delta\phi^I = \phi^I-\bar{\phi}^I = Q^I - \dfrac{1}{2}\,\Gamma^I{}_{JK}\,Q^J Q^K  + \cdots\,.
    \label{inflationary perturbation}
\enq

Since both $\delta\phi^I$ and $Q^I$ are field-space vectors, we can decompose them into their adiabatic component $Q_\sigma$ and isocurvature component $Q_s$ as $Q^I = Q_\sigma T^I+ Q_s N^I$, where a priori both components are dynamic. On the other hand, the study of spacetime perturbations and their mixing with $Q^I$ is simplified by using the ADM form \cite{Arnowitt2008, Maldacena2003} of the metric \eqref{FLRW metric}:

\beq
\dif s^2 = -N^2\dif \phi^2 + \gamma_{ij}\,(N^i\dif \phi+\dif x^i)(N^j\dif \phi + \dif x^j)\,,
\label{ADM metric}
\enq

where $\dif \phi$ is a time parameterization, $N$ is the lapse function, $N^i$ is the shift vector, and $\gamma_{ij}$ is the induced 3-metric on the hypersurface $\Sigma_t$. The rationale behind the ADM formalism in the study of cosmological perturbations is to expand the action \eqref{accion marco einstein} to second order in a power series of the perturbations. This approach was introduced by Maldacena \cite{Maldacena2003} for single-field inflation models. Since $N$ and $N^i$ are non-dynamical Lagrange multipliers, the 3-metric $\gamma_{ij}$ contains the physical information of the system. By fixing a time parameterization and choosing the shift vector, we fix $N$ and $N^i$, which is achieved by choosing a gauge. In inflation and perturbation theory, it is common practice to employ the spatially flat gauge, where $\gamma_{ij}^\mathrm{flat} = a^2(t)\delta_{ij}$, and all dynamics are contained within the perturbation $\delta \phi^I$. This gauge has been widely used in the literature \cite{Sebastian2012, Bartolo2001, Wands2002, LANGLOIS2003, Langlois2008} and has the advantage that the components of the perturbation $\delta\phi^I$ are Mukhanov-Sasaki (MS) variables, thus satisfying an MS-type equation \cite{Sasaki1996, MUKHANOV1998, Wands2002, Gordon2000}. However, in this work, we choose to work with the comoving gauge, defined as \cite{Saenz2020, Seery2005, Lyth2005, Lyth2005a}:

\beq
\gamma_{ij}^\mathrm{com} = a^2(t)e^{2\mathcal{R}}\delta_{ij}\,,\quad T_I Q^I = 0\,.
\label{3-metric}
\enq

where $\mathcal{R}$ is the gauge-invariant comoving curvature perturbation \cite{MUKHANOV1992}. The main feature of this gauge is that the adiabatic component of $Q^I$ vanishes, but the isocurvature part is $Q_s = N_I Q^I$. Consequently, in the comoving gauge, the perturbations are $\mathcal{R}$ and $Q_s$. The action at second order in these variables is written as

\begin{multline}
     S^{(2)} = \int \dif^4 x\, a^3\left[ \epsilon\left(\dot{\mathcal{R}}^2-a^{-2}\delta^{ij}\partial_i\mathcal{R}\partial_j\mathcal{R}\right) + \dfrac{1}{2}\left(\dot{Q}_s^2-a^{-2}\delta^{ij}\partial_iQ_s \partial_j Q_s- m_\mathrm{iso}^2Q_s^2\right)\right. \\
     \left.+ 2H \eta_\perp \sqrt{2\epsilon}\, Q_s \dot{\mathcal{R}} \right]\,,
        \label{second order action}
\end{multline}

where $\epsilon$ is the slow-roll parameter given in \eqref{SR parameters}, while $m^2_\text{iso}$ is the effective mass of the isocurvature mode $Q_s$, given by

\beq
    m_\mathrm{iso}^2 = N^I N^J \nabla_I V_J + H^2\epsilon\,R_\text{fs}-(H\eta_\perp)^2\,,
    \label{iso mass}
\enq

where $R_\text{fs} = -1/3$ is the field-space Ricci scalar of the model in units of $M_p$, defined via $G_{IJ}$ \eqref{field space metric}. The first term $V_{NN} = N^I N^J \nabla_I V_J$ represents the usual mass contribution given the potential \eqref{potential Einstein}:

\begin{align}
   \nonumber V_{NN} &= N^I N^J \nabla_I V_J = N^I N^J(\partial_I \partial_J V - \Gamma^K{}_{IJ}\partial_K V) \\[0.3cm]
     &= \dfrac{1}{\dot{\sigma}^2}\left(e^{\alpha\phi}\dot{\phi}^2 V_{hh} + e^{-\alpha \phi}\dot{h}^2 V_{\phi\phi} - 2\dot{\phi}\,\dot{h}V_{h\phi} - \dfrac{\alpha}{2}(\dot{\phi}^2 V_\phi + 2\dot{\phi}\,\dot{h}V_h ) \right)\,,
    \label{usual mass term}
\end{align}

which depends on the derivatives of the potential \eqref{potential Einstein} and the background dynamics. However, this can be considerably simplified if we take into account the valley-approach \eqref{valley} and the results in \eqref{valley approach}. By explicitly substituting the form of $N^I$ and the potential, and taking \eqref{valley} into account, we obtain

\beq
    V_{NN} \simeq \dfrac{\xi_h(24\lambda\xi_s + \xi_h(1 + 6\xi_h))}{\lambda\xi_s}\,H^2\,.
\enq

The second term in \eqref{iso mass} represents a contribution from the curved geometry of the field space, while the third term yields a negative contribution to the mass and is a correction due to the turn of the classical trajectory induced by \eqref{turning rate}. We observe that even if the turning rate is zero, multi-field effects do not vanish, as the second term in \eqref{iso mass} arising from the field-space geometry is non-trivial. Under the approximation \eqref{valley approach}, the mass $m_\text{iso}^2$ becomes

\beq
    \dfrac{m_\text{iso}^2}{H^2} \simeq \dfrac{\xi_h(24\lambda\xi_s + \xi_h(1 + 6\xi_h))}{\lambda\xi_s} -\dfrac{\epsilon}{3} - \dfrac{\tilde{\xi}^2}{1 + \tilde{\xi}^2}\,,
    \label{iso mass valley}
\enq

where $\tilde{\xi}^2 = \xi_h/6(\xi_h^2 + 4\lambda\xi_s)$. We observe that in the valley-approach, the mass of the $Q_s$ mode takes a simplified form, and the ratio $m_\text{iso}^2/H^2$ is effectively constant. This facilitates the search for analytical solutions to the equation for $Q_s$. On the other hand, if $\xi_h \gg 1$, the isocurvature mode is heavy and the turning rate is negligible, allowing us to ignore the evolution of $Q_s$. If we take $\xi_h \sim \lambda\xi_s$, the value of $m_\text{iso}^2$ is close to $H^2$; consequently, the $Q_s$ mode becomes light and the turning rate becomes significant. In this scenario, the single-field approximation fails, and one must adopt a quasi-single field inflation (QSFI) scenario \cite{Chen_2010,chen2010prd}, or numerically solve the equations for the modes $\mathcal{R}_k$ and $Q_s$, going beyond the single-field slow-roll approximation \cite{Wang2017}.

\subsection{Evolution of adiabatic and isocurvature perturbations}

The equations of motion for the perturbations $\mathcal{R}$ and $Q_s$ are obtained by varying the action \eqref{second order action}. In terms of cosmic time $t$ and in Fourier space, these are given by:

\beq
    \ddot{\mathcal{R}}_k + (3 + \delta)H\dot{\mathcal{R}}_k  + \dfrac{k^2}{a^2}\mathcal{R}_k = -\dfrac{2 H \eta_\perp}{\sqrt{2\epsilon}} [\dot{Q_s} + (3-\xi_\perp - \eta_{||} )HQ_s]
    \label{R ec.}
\enq

\begin{figure}[t]
    \centering
    \includegraphics[width = 0.48 \textwidth]{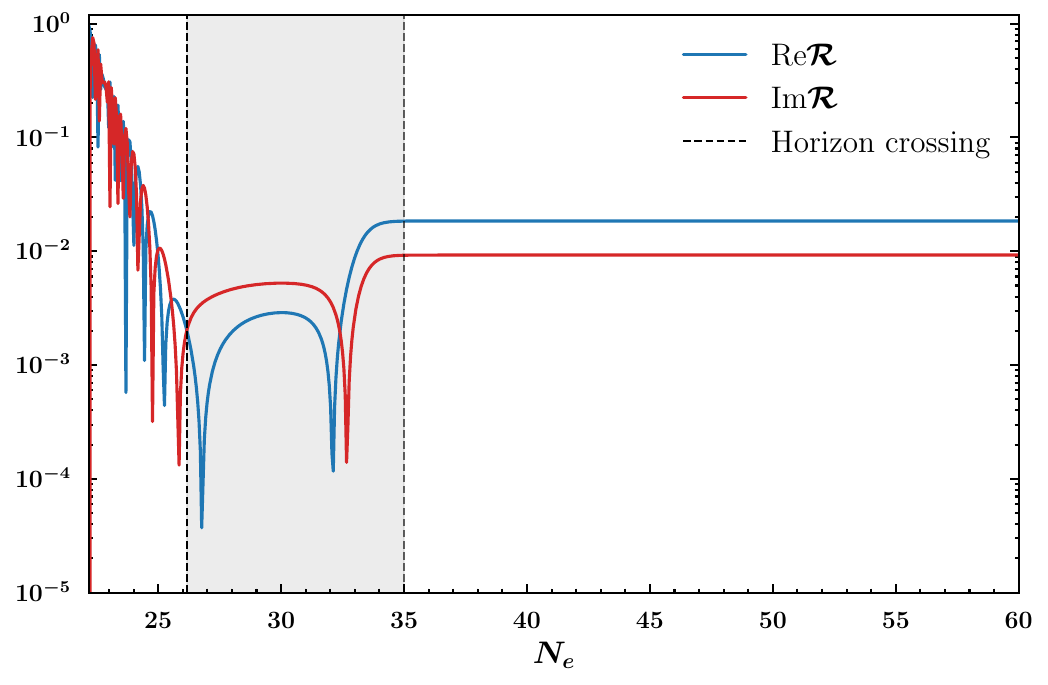}
    \includegraphics[width = 0.48 \textwidth]{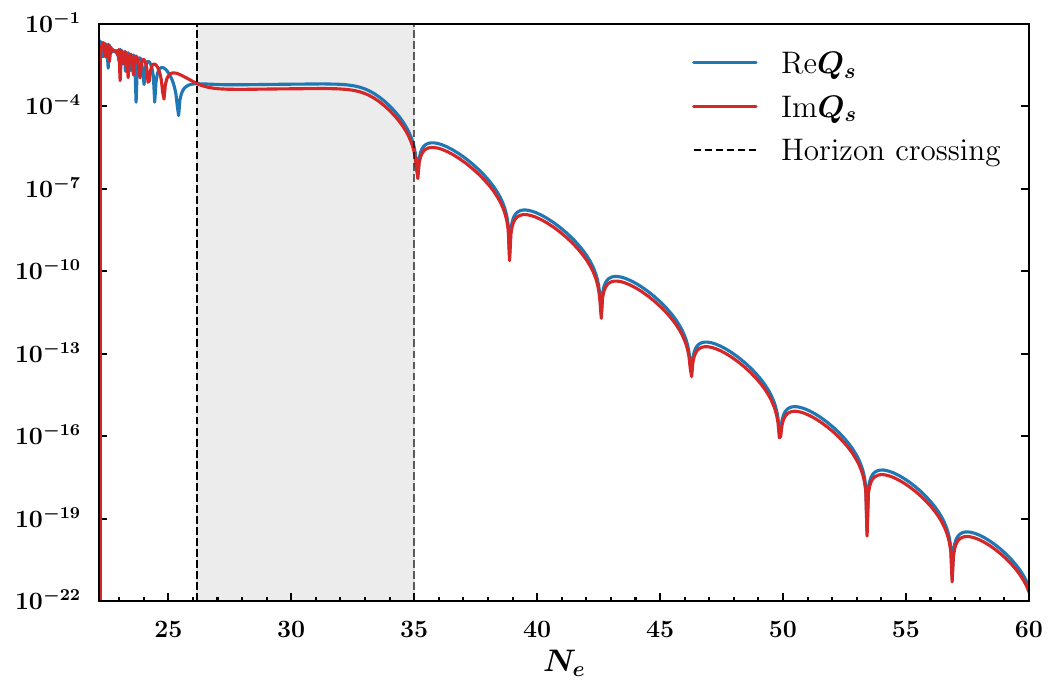} \\[0.3cm]
    \includegraphics[width = 0.48 \textwidth]{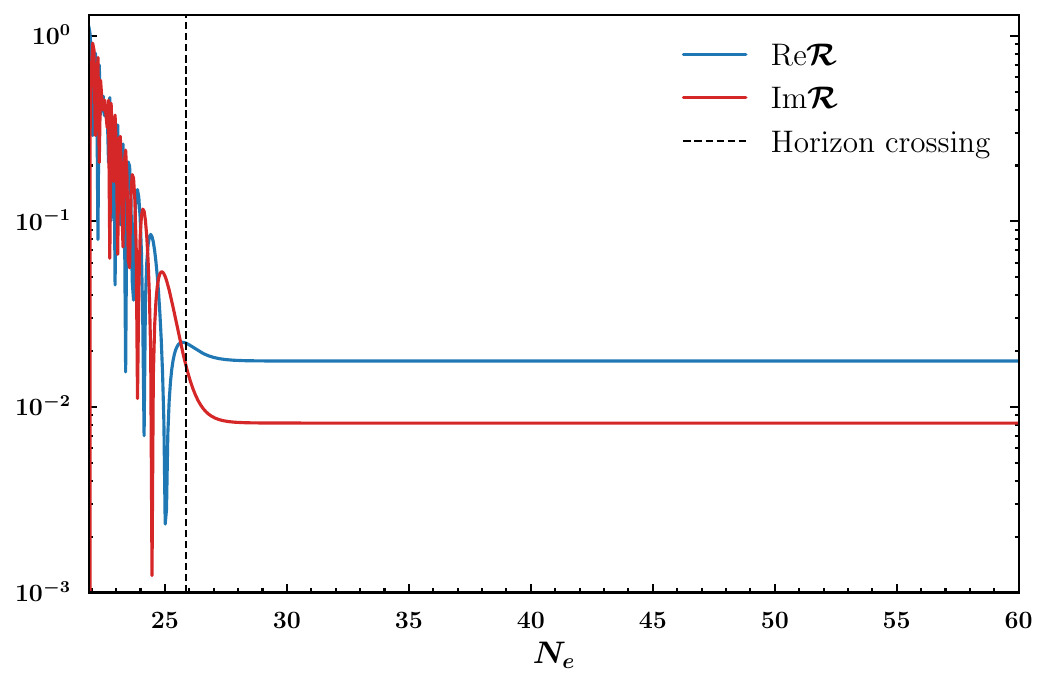}
    \quad
    \includegraphics[width = 0.48 \textwidth]{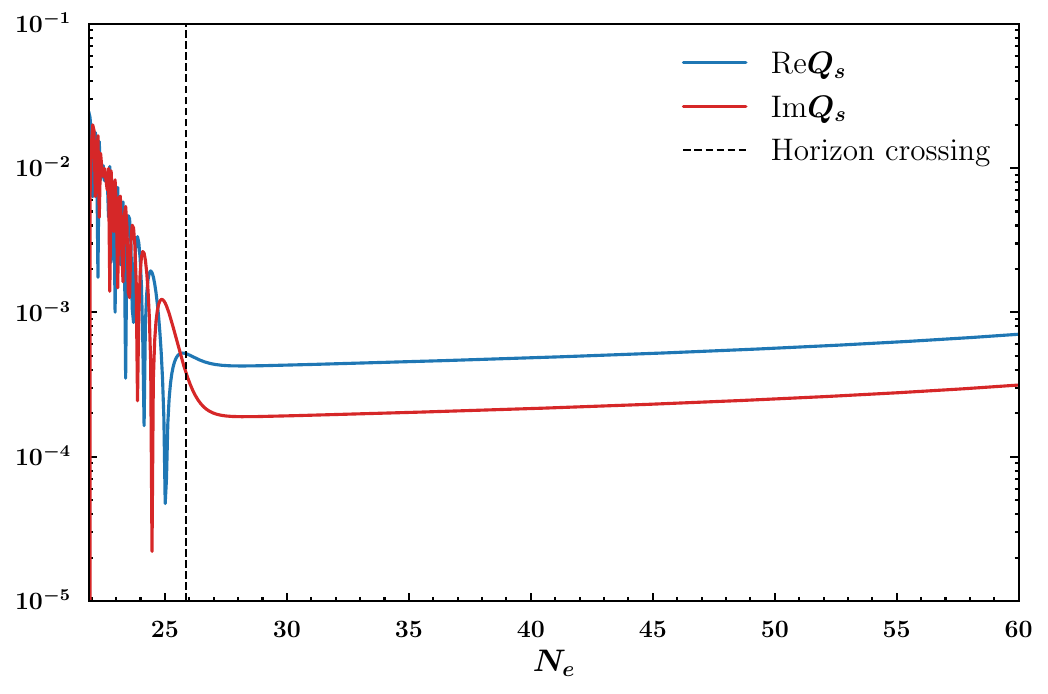}

    \caption{Numerical solution of equations \eqref{R equation efolds} and \eqref{Qs equation efolds} for the space parameter $\xi_h= 0.1$ (top panel) and $\xi_h= 10^{-9}$ (bottom panel) with $\lambda = 10^{-10}$ and $\xi_s = 4\times 10^8$ fixed. We also fixed the initial conditions to $\phi_0 = 5.7$ and $h_0 = 10^{-4}$ in Planck units. The vertical dashed line indicates horizon crossing for the pivot scale $k_* = 0.05\,\text{Mpc}^{-1}$.}
    \label{fig3}
\end{figure}

\beq
    \ddot{Q}_s + 3H\dot{Q}_s + \left(\dfrac{k^2}{a^2} + m_\mathrm{iso}^2\right) Q_s = 2 H \eta_\perp \sqrt{2\epsilon} \dot{\mathcal{R}}_k\,,
    \label{Qs ec.}
\enq

where $\delta = 2(\epsilon - \eta_{||})$ and $\xi_\perp = -\dot{\eta}_\perp/H\eta_\perp$. The action \eqref{second order action} and equations \eqref{R ec.} and \eqref{Qs ec.} are exact; that is, no approximations were made in their derivation. Generally speaking, we can integrate the modes $\mathcal{R}_k$ and $Q_s$ on super-Hubble scales ($k\ll aH$) for $\eta_\perp \neq 0$. The equation for the isocurvature mode $Q_s$ becomes

\beq
    \ddot{Q}_s + 3H\dot{Q}_s + m_\text{eff}^2 Q_s \simeq 0\,,
    \label{Qs SH evo}
\enq

where the effective mass is given by $m_\text{eff}^2 = m_\text{iso}^2 + 4(H\eta_\perp)^2$, with $m_\text{iso}^2$ defined in \eqref{iso mass}. In this case, the turning rate parameter always makes a positive contribution to the mass of the $Q_s$ mode, acting as a source for the adiabatic mode such that $\dot{\mathcal{R}_k}\simeq - \sqrt{2}H\eta_\perp Q_s/\sqrt{\epsilon}$. For $\eta_\perp = 0$, it is clear that the mode $\dot{\mathcal{R}}_k$ decays exponentially, and thus $\mathcal{R}_k$ converges to a constant value, recovering the standard single-field inflation result. 
It is worth noting that the mass of the $Q_s$ mode does not depend on the super-Hubble evolution of the adiabatic mode, so it can be calculated independently. Thus, the mass $m_\text{iso}^2$ is solely a function of the background variables and contains all the relevant physical information for $Q_s$. 

To visualize this behavior and track the full evolution of the $\mathcal{R}_k$ and $Q_s$ modes on super-Hubble scales, it is convenient to perform a numerical analysis of the perturbations. As with the background dynamics, it is more efficient to numerically integrate equations \eqref{R ec.} and \eqref{Qs ec.} in terms of $N_e$:

\beq
    \mathcal{R}''_k + \left(1 - \epsilon + 2\dfrac{\dif \log z}{\dif N} \right)\mathcal{R}'_k + \left(\dfrac{k}{aH}\right)^2 \mathcal{R}_k = -\dfrac{2\eta_\perp}{\sqrt{2\epsilon}}[Q_s' + (3-\xi_\perp - \eta_{||})Q_s]\,,
    \label{R equation efolds}
\enq

\beq
    Q_s'' + (3-\epsilon)Q_s' + \left[\left(\dfrac{k}{aH}\right)^2 + \dfrac{m_\text{iso}^2}{H^2} \right]Q_s = 2\sqrt{2\epsilon}\, \eta_\perp\mathcal{R}'_k\,,
    \label{Qs equation efolds}
\enq

where $z = a\sqrt{2\epsilon}$, with $\epsilon$ given by \eqref{efolds SR}. We impose the Bunch-Davies vacuum as the initial condition for the modes $\mathcal{R}_k$ and $Q_s$, taking into account that the Mukhanov-Sasaki variables $u_k$ and $v_k$ are defined as $u_k = z\mathcal{R}_k$ and $v_k = a Q_s$. One must ensure that for each mode, the perturbations are initially deep inside the horizon ($k \gg aH$). If $N_\text{hc}$ is the e-fold corresponding to the horizon crossing of the mode, $k = aH$, we start the evolution 5 e-folds earlier, i.e., $N_\text{ini} = N_\text{hc} - 5$. This guarantees that for each mode $k$, the perturbations are well inside the horizon at the start of the integration.
In terms of the number of e-folds, the initial conditions for $\mathcal{R}_k$ and $Q_s$ are given by

\beq
    \mathcal{R}_k(N_\text{ini}) = \dfrac{1}{z(N_\text{ini})\sqrt{2k}}\,,\quad Q_s(N_\text{ini}) = \dfrac{1}{a_0\sqrt{2k}}\,,
    \label{BD vacuum}
\enq

where $a_0 = a(N_\text{ini})$ is the value of the scale factor at $N_\text{ini}$. We separate the real and imaginary parts of equations \eqref{R equation efolds} and \eqref{Qs equation efolds}, as well as the initial conditions \eqref{BD vacuum}, to achieve greater numerical stability in the solution. In Fig.~\ref{fig3}, we display the evolution of the perturbations $\mathcal{R}_k$ and $Q_s$ as a function of $N_e$ for the pivot scale $k_* = 0.05\,\text{Mpc}^{-1}$ and for two  space parameters $\xi_h = 0.1$ and $\xi_h = 10^{-9}$. As observed for $\xi_h = 0.1$, the mode $\mathcal{R}_k$ continues to evolve after horizon crossing due to the turning rate $\eta_\perp$ in field space, until it eventually freezes out. This evolution is driven by the coupling between the $\mathcal{R}_k$ and $Q_s$ modes via the turning rate parameter $\eta_\perp$. This transfer from the $Q_s$ mode is consistent with the previous discussion. In contrast, the isocurvature mode $Q_s$ evolves after crossing and is exponentially suppressed as a consequence of its effective mass $m_\text{iso}^2 \gtrsim H^2$, limiting its late-time contribution as a source for $\mathcal{R}_k$.
For the case $\xi_h = 10^{-9}$, the situation is identical to the single-field case: the adiabatic perturbation  freezes out after horizon crossing since there is no coupling with isocurvature mode $Q_s$.

\section{Primordial power spectra}
\label{sec4}

We have mentioned that the background behavior shown in Fig.~\ref{Fig2} triggers multi-field effects, which are reflected in the evolution of the primordial perturbations $\mathcal{R}_k$ and $Q_s$ on super-Hubble scales (see Fig.~\ref{fig3}). In particular, these effects can generate additional features in the primordial spectrum, depending on the value of the constant coupling $\xi_h$.  Although the variables $\mathcal{R}_k$ and $Q_s$ are convenient for integrating the perturbation equations, it is more useful to express the results in terms of the dimensionless isocurvature perturbation

\begin{equation}
    \mathcal{S}_k \equiv \frac{H}{\dot{\sigma}}\, Q_s\,,
\end{equation}

which allows for a direct comparison with observational literature. Super-Hubble evolution induces a correlation between the adiabatic mode $\mathcal{R}_k$ and the isocurvature mode $\mathcal{S}_k$. Therefore, the primordial spectrum is characterized by three functions:

\begin{equation}
 \mathcal{P}_{\mathcal R}(k) = \frac{k^3}{2\pi^2}\,|\mathcal{R}_k|^2, \qquad
 \mathcal{P}_{\mathcal S}(k) = \frac{k^3}{2\pi^2}\,|\mathcal{S}_k|^2, \qquad
 C_{\mathcal{RS}}(k) = \frac{k^3}{2\pi^2}\,{\rm Re}\!\left(\mathcal{R}_k \mathcal{S}_k^\ast\right).
 \label{powerspectrum}
\end{equation}

The quantity $C_{\mathcal{RS}}(k)$ is known as the \emph{cross-correlation spectrum} and quantifies the degree of correlation between both modes. From the adiabatic and isocurvature spectra, we can define the relative isocurvature fraction

\begin{equation}
    \beta_{\rm iso}(k) =
    \frac{\mathcal{P}_{\mathcal S}(k)}
         {\mathcal{P}_{\mathcal R}(k) + \mathcal{P}_{\mathcal S}(k)}\,,
    \label{beta_iso}
\end{equation}

which can be compared with the most recent observational limits from Planck~\cite{planck2018x}. Additionally, the correlation between modes is parameterized via the correlation angle

\begin{equation}
    \cos\Delta =
    \frac{C_{\mathcal{RS}}}
         {\sqrt{\mathcal{P}_{\mathcal R}\,\mathcal{P}_{\mathcal S}}}\,,
         \label{croos angle}
\end{equation}

which takes values in the interval $(-1,1)$ for each mode $k$. The presence of a non-zero cross-spectrum is a purely multi-field feature: the transfer $\mathcal{S}_k\!\to\!\mathcal{R}_k$ on super-Hubble scales is controlled by the turning rate parameter $\eta_\perp$, as follows from the evolution equation for $\dot{\mathcal{R}}_k$. Therefore, both the amplitude and shape of $C_{\mathcal{RS}}$ act as direct tracers of the field-space trajectory geometry.  Observationally, a non-zero value of $C_{\mathcal{RS}}$ implies that the primordial spectrum cannot be described solely by the adiabatic and isocurvature amplitudes, but requires a relative phase between the modes. This effect can modify the CMB angular power spectra in specific ways through linear interference between the adiabatic and isocurvature terms, an aspect we will discuss in Sec.~\ref{sec5} using the obtained primordial spectrum as input for Boltzmann codes such as \texttt{CLASS} \cite{class} \footnote{The public code can be found in the repository \url{https://github.com/lesgourg/class_public.git}.}.

\subsection{Numerical results}

Based on the numerical solutions for the background (Fig.~\ref{Fig2}) and perturbations (Fig.~\ref{fig4}), we evaluate the primordial spectrum at the end of inflation, defined by the e-fold $N_{\rm end}$ such that $\epsilon_H(N_{\rm end}) = 1$. For each mode $k$, we identify the horizon crossing time $N_{\rm hc}$ via the condition $k = aH$, while $N_{\rm CMB}$ denotes the e-fold corresponding to the pivot scale $k_* = 0.05\,\mathrm{Mpc}^{-1}$, which we fix it in $N_{\rm CMB} = 50$ for all the cases. In this way, each $k$ value is uniquely associated with a crossing instant during the inflationary evolution. To encompass both the observable CMB scales and the modes exiting the horizon in the final stages of inflation, we construct a logarithmic grid of 1000 $k$ values in the interval $[k_{\rm min},k_{\rm max}]$, where $k_{\rm min} = aH(N_{\rm hc}-7)$ and $k_{\rm max} = aH(N_{\rm end}-4)$.
This range covers from $k\sim 10^{-6}\,\mathrm{Mpc}^{-1}$ to $k\sim 10^{19}\,\mathrm{Mpc}^{-1}$ and ensures that the power spectrum calculation incorporates modes experiencing the $\mathcal{S}_k\!\rightarrow\!\mathcal{R}_k$ transfer throughout all dynamic stages of the model. We evaluate Eq. \eqref{powerspectrum} at the end of inflation for each mode $k$; unlike the single-field case, it is not evaluated at horizon crossing because the power spectrum undergoes super-Hubble evolution for the parameter space of interest. This guarantees that the perturbation $\mathcal{R}_k$ has frozen out and that $Q_s$ exhibits the expected exponential decay.
Our approach covers two important cases: $\xi_h\ll 1$ and $\xi_h \sim \mathcal{O}(0.1)$; in both instances, we fix $\lambda = 10^{-10}$ and $\xi_s = 4\times 10^{8}$. The initial condition is also set to $h_0 = 10^{-4} M_p$. Our results in this paper for the background, primoridal perturbations and the power spectrum can be produced by our numerical code which is available in \cite{pineda_data_2026}.

\subsubsection{Case $\xi_h \ll 1$}
\label{iso remain}

\begin{figure}[t]
    \centering
    \includegraphics[width = 0.48 \textwidth]{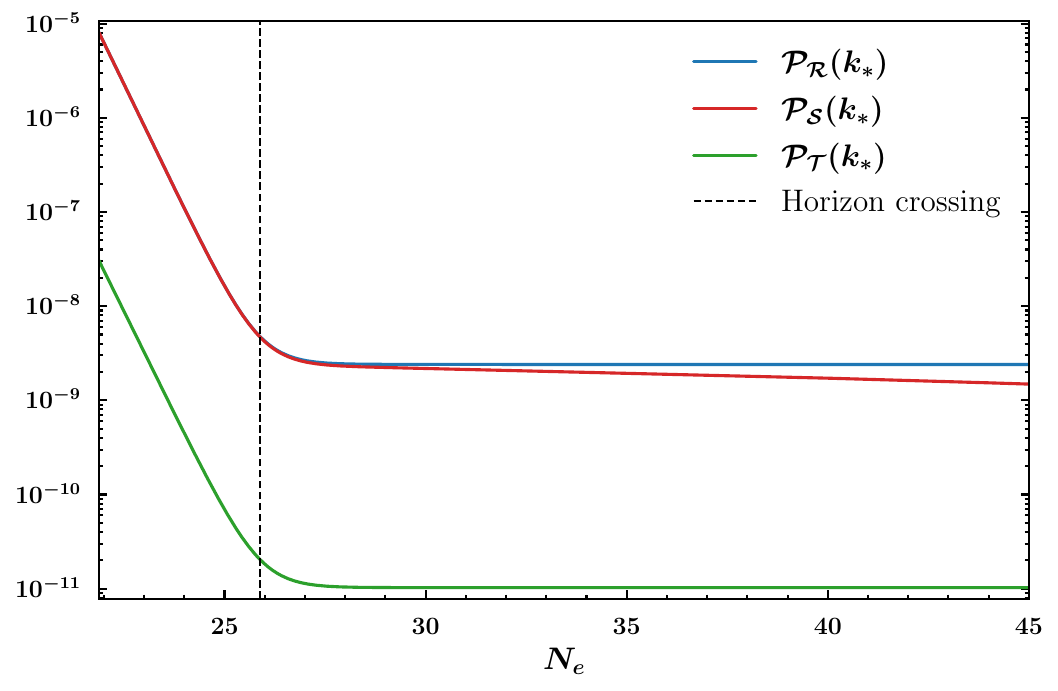}\quad
     \includegraphics[width = 0.48 \textwidth]{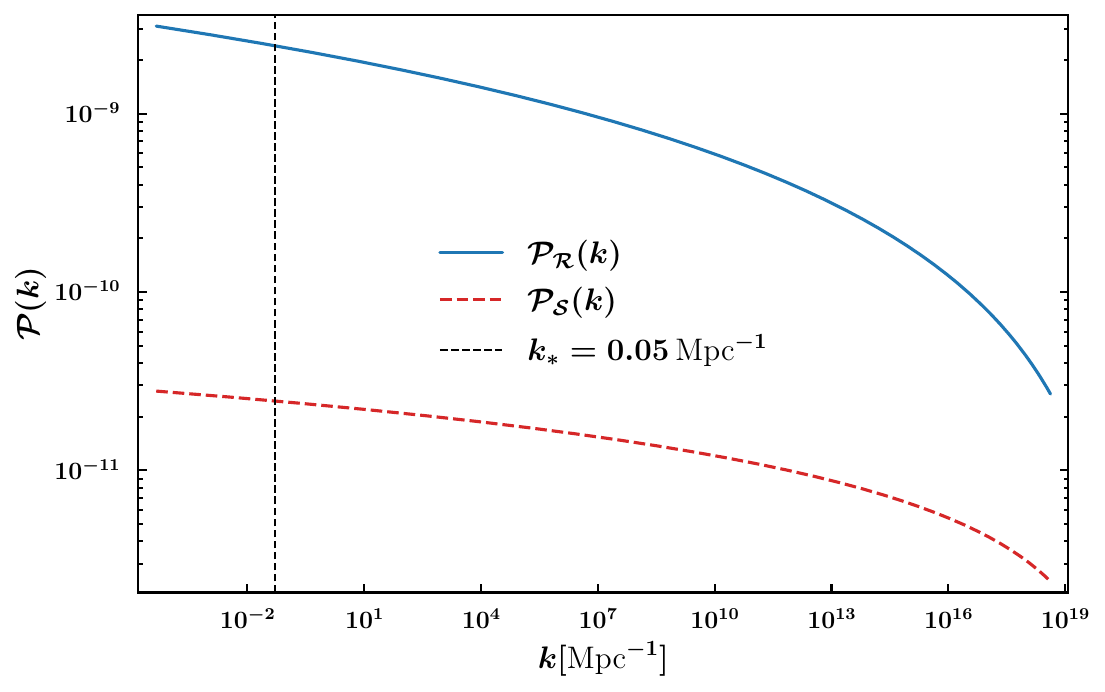}\vspace{0.2cm}
     \includegraphics[width = 0.48 \textwidth]{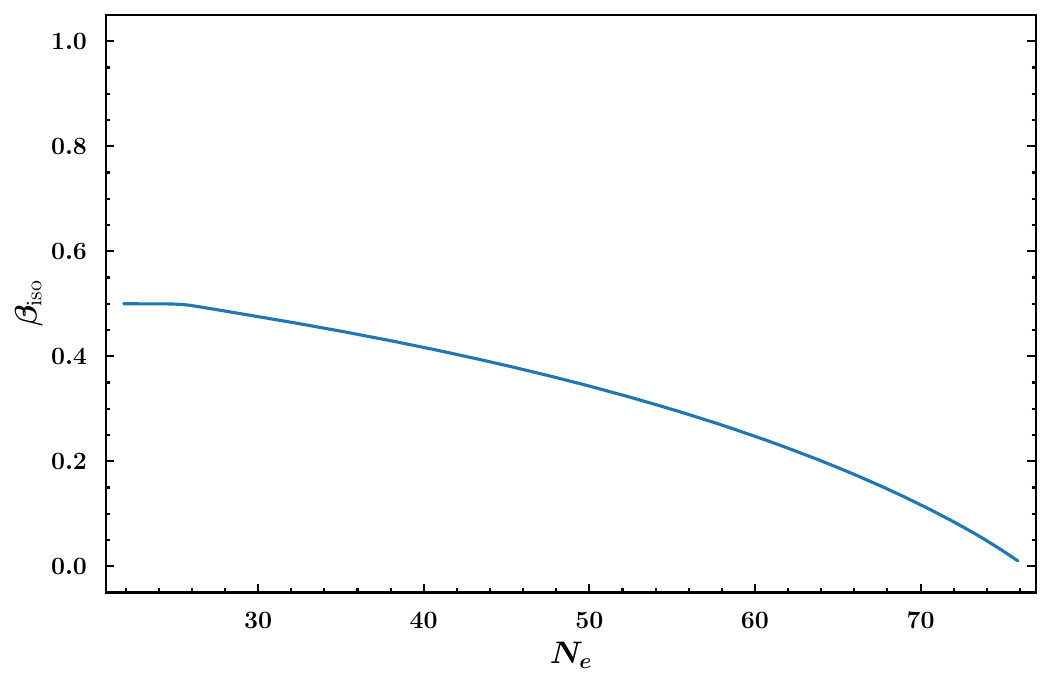}\vspace{0.2cm}
     \includegraphics[width = 0.48 \textwidth]{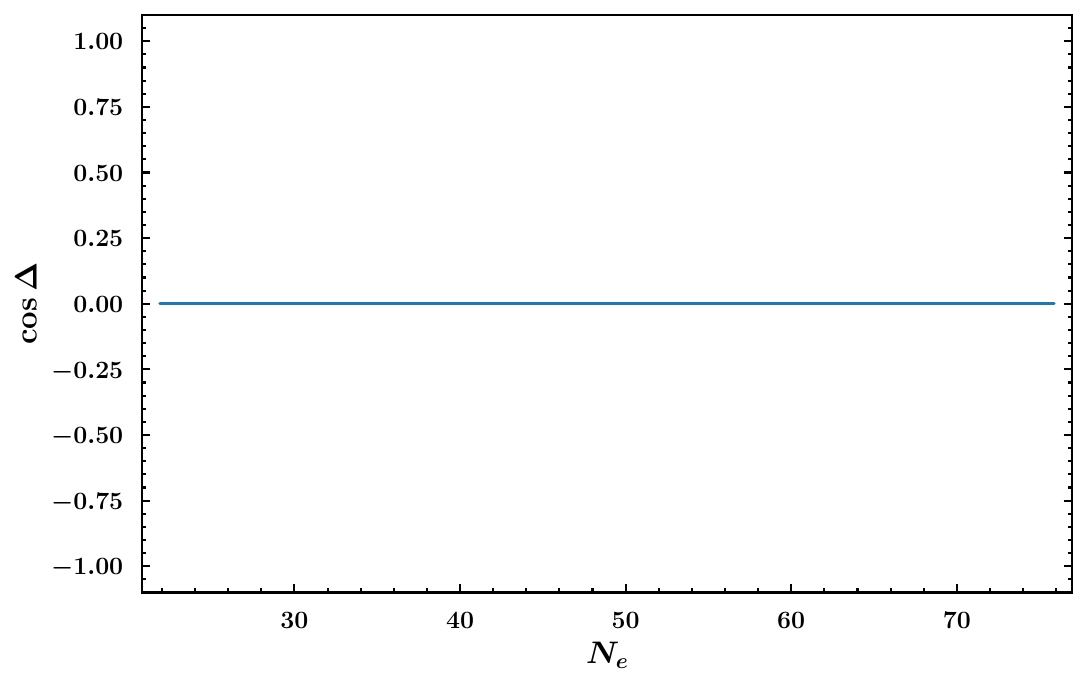}
    \caption{Primordial power spectra in the weak-coupling regime $\xi_h = 10^{-9}$. \textbf{Left:} Time evolution of the adiabatic ($\mathcal{P}_{\mathcal{R}}$), isocurvature ($\mathcal{P}_{\mathcal{S}}$), and tensor ($\mathcal{P}_{\mathcal{T}}$) power spectra evaluated at the pivot scale $k_* = 0.05 \text{ Mpc}^{-1}$ (top), and the isocurvature fraction $\beta_{\text{iso}}$ (bottom) as a function of the number of e-folds $N_e$.  \textbf{Right:} The resulting primordial power spectra for the adiabatic and isocurvature modes as a function of the comoving wavenumber $k$, and cross-correlation $\cos \Delta$ (bottom) illustrating a featureless adiabatic spectrum and persistent isocurvature power at the end of inflation.}
    \label{fig4}
\end{figure}

This scenario simplifies both the isocurvature mass $m_\text{iso}^2$ and the turning rate $\eta_\perp$ within one of the valleys. As discussed in Ref.~\cite{GUNDHI2020}, the potential loses its two-valley structure, which degenerates into a single valley centered at $h = 0$. Consequently, the dynamics may commence within the valley or precisely on the ridge, from where the field eventually oscillates and falls into the valley towards the potential minimum at $(0,0)$. Within the valley, the turning rate vanishes ($\eta_\perp \to 0$) for $\xi_h \ll 1$, as shown in Appendix \ref{valley approach}, while the mass of the $Q_s$ mode is given by $m_\text{iso}^2/H^2 \simeq -\epsilon/3$ in accordance with Eq. \eqref{iso mass valley}. Thus, the mass remains light and negative during inflation. This causes the $Q_s$ mode to grow on super-Hubble scales ($k \ll aH$). On the other hand, the fact that $\eta_\perp \simeq 0$ in this parameter space renders the perturbations statistically independent as long as this condition holds, since the generation of correlations and the energy transfer from the isocurvature to the curvature mode require the perturbations to be coupled via $\eta_\perp$. However, isocurvature modes are produced and survive until the end of inflation, owing to the lightness of the $Q_s$ field. Consequently, both $\mathcal{R}_k$ and $\mathcal{S}_k$ evolve independently throughout the entire inflationary period. This behavior is displayed in Fig.~\ref{fig4}. The top-left panel displays the evolution of the primordial spectra $\mathcal{P}_\mathcal{R}(k)$, $\mathcal{P}_\mathcal{S}(k)$, and $\mathcal{P}_\mathcal{T}(k)$ as a function of the number of e-folds $N_e$, evaluated at the pivot scale $k_* = 0.05\,\text{Mpc}^{-1}$. At horizon crossing, the power spectrum of $\mathcal{R}_k$ freezes out as expected (since $\eta_\perp \simeq 0$), but $\mathcal{P}_\mathcal{S}(k)$ is not suppressed at the end of inflation, and its contribution remains non-zero. This is because the effective mass of the $Q_s$ mode remains light throughout the evolution, in agreement with our analysis. The tensor spectrum remains practically unaltered, as there is no coupling between scalar and tensor degrees of freedom at linear order in perturbations.
The top-right panel shows the primordial power spectrum at the end of inflation. As observed, the curvature power spectrum $\mathcal{P}_\mathcal{R}(k)$ exhibits no features; at large scales, its amplitude is of the order of $10^{-9}$, while at small scales, it presents a suppression in magnitude. On the other hand, the isocurvature power spectrum is of the order of $10^{-11}$. The bottom panels display the isocurvature fraction $\beta_{\rm iso}$ (left) and the cross-correlation $\cos\Delta$ (right) as a function of e-folds. The cross-correlation vanishes because the turning rate is negligible in this parameter space, but the isocurvature fraction is non-zero at the end of inflation, reaching a value of $\beta_{\rm iso} \approx 0.0101$ at the pivot scale $k_* = 0.05\,{\rm Mpc}^{-1}$. 
This value corresponds to an isocurvature contribution of the order of $1\%$, compatible with current limits from the Planck analysis on CDM isocurvature models $\beta_{\rm iso}(k_*) < 0.038$ at $95\%)$ CL \cite{planck2018x}, although a precise comparison depends on the correlation and the spectral index of the isocurvature component. The absence of correlation simplifies the impact on the angular spectra, implying that the applicable observational bounds correspond to the uncorrelated Planck case. Furthermore, we verify that the inclusion of this fraction in \texttt{CLASS} does not generate significant discrepancies with the observed angular spectrum (see Section \ref{sec5}).

\begin{figure}[t!]
     \centering
    \includegraphics[width = 0.48 \textwidth]{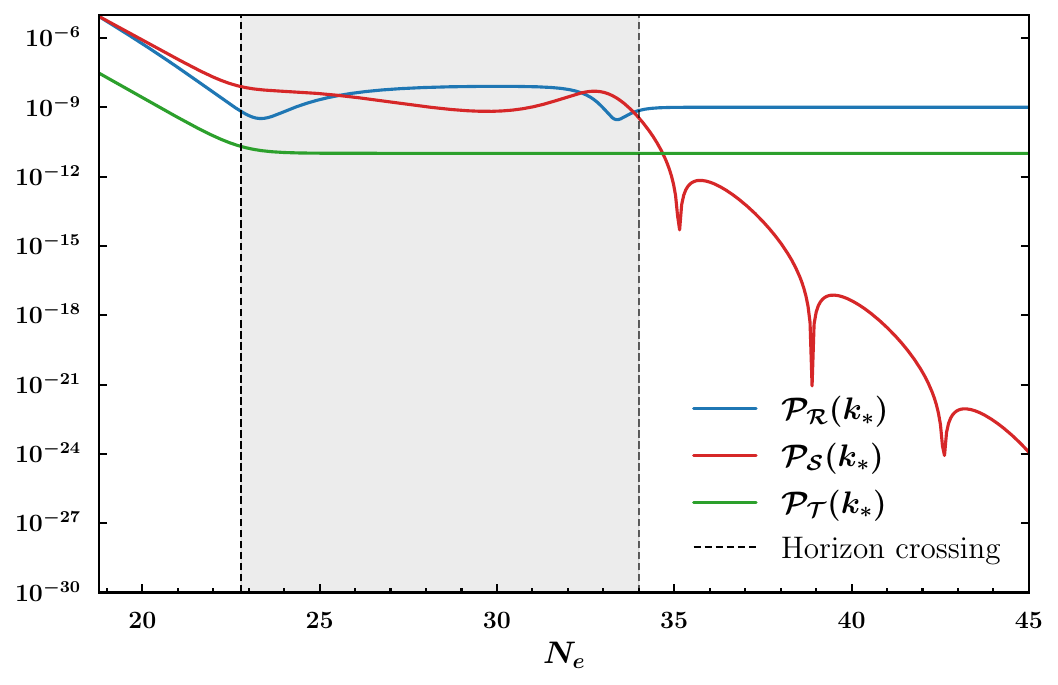}\quad
     \includegraphics[width = 0.48 \textwidth]{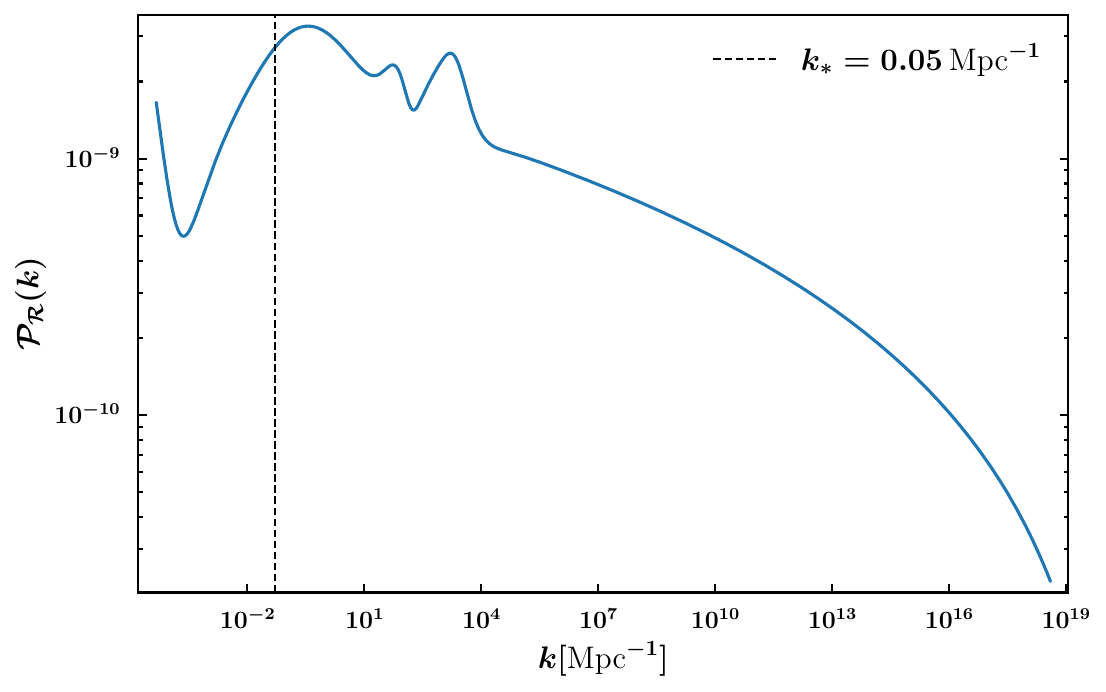}\vspace{0.2cm}
     \includegraphics[width = 0.48 \textwidth]{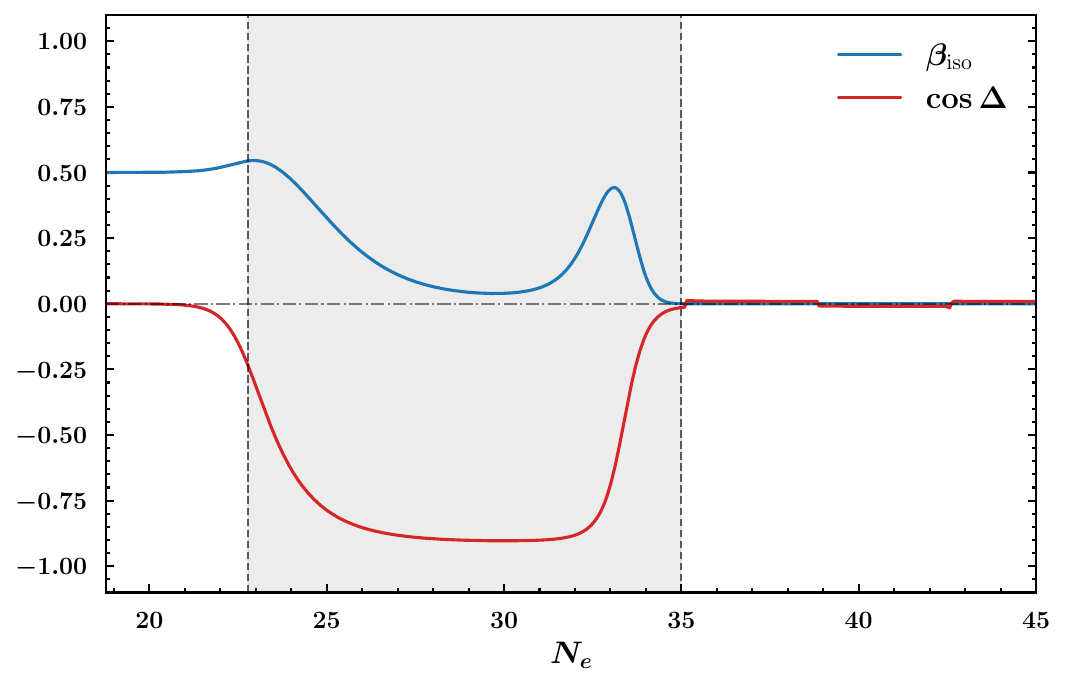}\vspace{0.2cm}
     \includegraphics[width = 0.48 \textwidth]{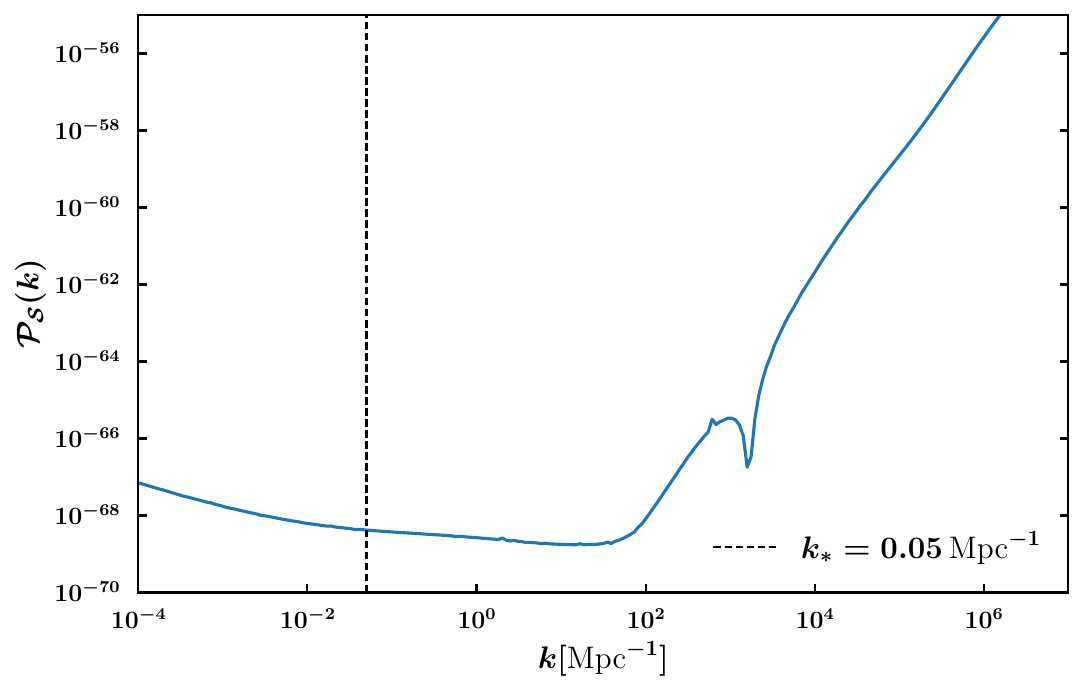}

    \caption{Evolution of the primordial perturbations for the benchmark Higgs-$R^2$ model $\xi_h = 0.1$. \textbf{Left:} Time evolution of the adiabatic ($\mathcal{P}_{\mathcal{R}}$), isocurvature ($\mathcal{P}_{\mathcal{S}}$), and tensor ($\mathcal{P}_{\mathcal{T}}$) power spectra evaluated at the pivot scale $k_* = 0.05 \text{ Mpc}^{-1}$ (top), alongside the isocurvature fraction $\beta_{\text{iso}}$ and cross-correlation $\cos \Delta$ (bottom) as a function of the number of e-folds $N_e$. The shaded area indicates the transient turning phase in the hyperbolic field space. \textbf{Right:} The resulting primordial power spectra for the adiabatic (top) and isocurvature (bottom) modes as a function of the comoving wavenumber $k$, evaluated at the end of inflation.}
    \label{fig5}
\end{figure}

\subsubsection{Case $\xi_h \sim \mathcal{O}(10^{-1})$}

For the parameter space where $\xi_h \sim 0.1$, we observe a distinct phenomenological behavior compared to the small coupling limit. We find that the isocurvature component is strongly anti-correlated with the adiabatic mode at the moment of horizon crossing for the pivot scale. However, $\mathcal{P}_{\mathcal{S}}$ is fully suppressed during super-Hubble evolution, yielding a vanishing isocurvature fraction at $N_{\rm end}$ within numerical precision. The strong anti-correlation at crossing acts as a destructive interference mechanism that suppresses the adiabatic power at specific scales, imprinting a feature in the primordial spectrum without leaving a relic isocurvature contribution at the end of inflation. The top-left panel of Fig.~\ref{fig5} illustrates the spectral evolution at the pivot scale $k_* = 0.05\,{\rm  Mpc}^{-1}$. The difference with respect to the $\xi_h = 10^{-9}$ case is striking: following horizon crossing, the curvature power spectrum continues to undergo significant evolution driven by the turning rate $\eta_\perp$ until it stabilizes at a constant value, marking its freeze-out. Conversely, the isocurvature spectrum is exponentially suppressed, reflecting its large effective mass and reducing its late-time impact on $\mathcal{R}_k$. This dynamics ensures that the isocurvature fraction $\beta_{\rm iso}(N_\text{end})$ becomes negligible. 

The tensor spectrum remains practically unaltered in both scenarios, as expected. The top-right panel displays the resulting curvature power spectrum. Distinct oscillatory features arise naturally at  scales $k \lesssim 10^4\,{\rm Mpc}^{-1}$, corresponding to the transition phase where the turning rate is significant and the slow-roll parameter exhibits a transient bump. At smaller scales, the spectrum recovers the scale-invariant prediction typical of single-field inflation driven by the scalaron ($R^2$) or the Higgs field $h$. The notable difference in this regime is that the amplitude of the adiabatic mode has been shifted due to the energy transfer from the isocurvature mode during the turn in field space. The bottom-left panel reveal a transient phase of coupling induced by the turning rate $\eta_\perp$ around $N_e \simeq 22-35$. This interaction generates a significant cross-correlation, reaching values of $\cos \Delta \approx -0.786$ at $k_* = 0.05\,{\rm Mpc}^{-1}$, and facilitates power transfer to the curvature spectrum, which is responsible for generating the features. Subsequently, due to the large effective mass of the $Q_s$ field, isocurvature perturbations decay exponentially, and the modes decouple completely before the end of inflation. This results in purely adiabatic final observables, i.e., $\beta_\text{iso} \to 0$, as shown in the same panel. Lastly, in the bottom-right panel we display the isocurvature power spectrum at the end of inflation. One can see that the isocurvature does not survive at the end of inflation, which is agreement with the isocurvature fraction $\beta_{\rm iso}$. 

\begin{figure}[t!]
    \centering
    \includegraphics[width =  \textwidth]{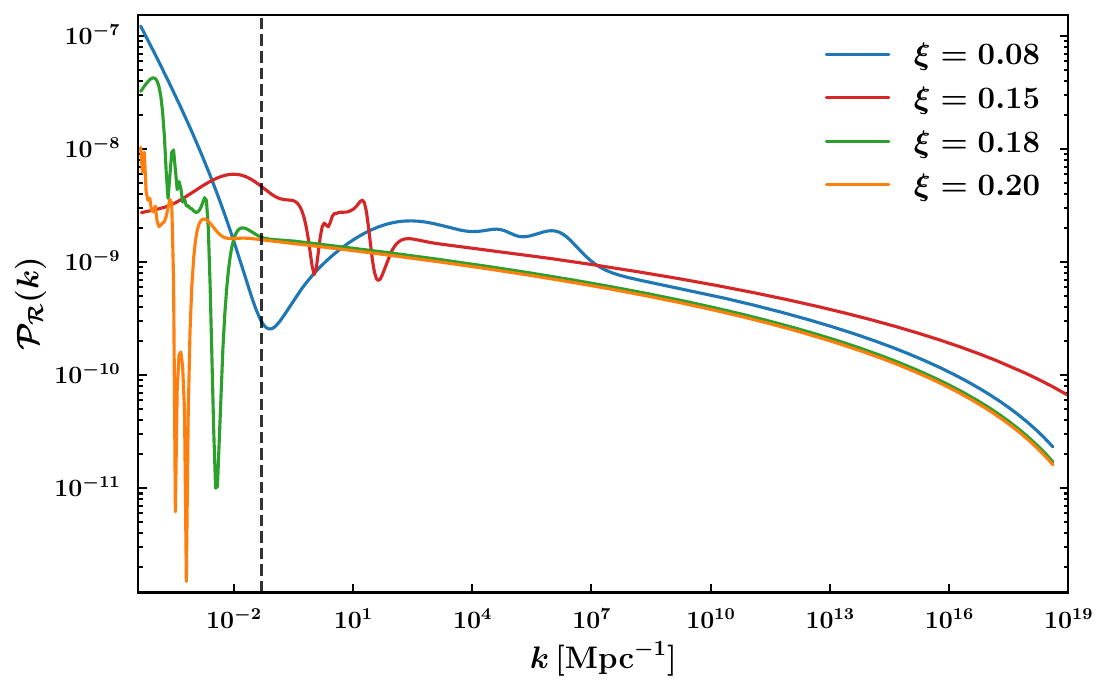}

    \caption{Adiabatic primordial power spectrum $\mathcal{P}_{\mathcal R}(k)$ evaluated at the end of inflation for $\lambda$ and $\xi_s$ fixed, and $\xi_h$ is varied over four values. The vertical dashed line corresponds to the scale pivot $k_* = 0.05\,{\rm Mpc}^{-1}.$}
    \label{fig6}
\end{figure}

For completeness, we performed a numerical scan over the non-minimal coupling in the range $\xi_h \in [0.08, 0.2]$. As illustrated in Fig.~\ref{fig6}, the position and amplitude of the resulting spectral features depend sensitively on $\xi_h$. Values close to $\xi_h \sim \mathcal{O}(0.1)$ lead to features affecting large and intermediate scales, while departures from this range shift the imprints to smaller scales or suppress them altogether.  This illustrates how transient multi-field dynamics can impact different CMB multipole ranges, depending on the underlying model parameters.

Besides the isocurvature fraction  \eqref{beta_iso} and the cross-correlation fraction \eqref{croos angle}, we also define the tensor to scalar ratio $r$ as the ratio between the amplitude of tensor and scalar perturbations, where scalar perturbations include adiabatic and isocurvature modes

\beq    
    r = \dfrac{\mathcal{P}_\mathcal{T}(k_*)}{\mathcal{P}_\mathcal{R}(k_*) + \mathcal{P}_\mathcal{S}(k_*)}
\enq 

which is evaluated at the pivot scale $k_* = 0.05 {\rm Mpc}^{-1}$. This expression can be rewritten in terms of the cross-correlation fraction as \cite{Bartolo2001, Wands2002,Dimarco2003,DiMarco2005, Braglia2020}

\beq  
r = -n_\mathcal{T}\left(1 - \dfrac{\mathcal{C}_{\mathcal{RS}}}{\mathcal{P}_\mathcal{R}\mathcal{P}_\mathcal{S}}\right) = -8n_\mathcal{T}\sin^2 \Delta\,,
\label{tensor_to_scalar}
\enq 

where $n_\mathcal{T}$ is the tensor spectral index, define as 

\beq
n_\mathcal{T} = \dfrac{\dif \log \mathcal{P}_\mathcal{T}}{\dif \log k}\,,
\enq

and the scalar spectral index $n_s$ is defined in the usual way 

\beq
    n_s - 1 = \dfrac{\dif \log \mathcal{P}_\mathcal{R}}{\dif \log k}\,.
\enq

The consistency relation \eqref{tensor_to_scalar} indicates that the amplitude of primordial gravitational waves is suppressed in multi-field inflation compared to the single-field expectation.

Current observational constraints from the combination of Planck, BICEP/Keck, and BAO data place an upper limit on the tensor-to-scalar ratio of $r < 0.036$ at $95\%$ CL \cite{BICEP_Keck_2021, planck2018x}, while the updated value reported by ACT + Planck + SPT + DESI for the scalar spectral index is $n_s \approx 0.9728 \pm 0.0027$ \cite{ACT_2025}. In our setup, we find that the Higgs--$R^2$ model predicts a tensor-to-scalar ratio of order $r \sim \mathcal{O}(10^{-3})$, which comfortably satisfies current observational bounds. The corresponding values of the main CMB observables are summarized in Table~\ref{table1}.

\subsection{Hyperbolic field-space geometry and isocurvature modes}

As we have seen above, the field-space geometry is hyperbolic with a scalar curvature given by $ R_\text{fs} = - 1/3$ in Planck units. The second term of the isocurvature mass \eqref{iso mass} is a pure geometrical contribution that acts in the direction of rendering the effective isocurvature mass tachyonic if the geometry is hyperbolic \cite{Renaux_Petel_2016}. This phenomenon is called geometrical destabilization, and is a general mechanism by which heavy fields can be destabilized during inflation, i.e.,  the curvature of the field space manifold can dominate the stabilizing force
from the potential and destabilize inflationary trajectories. In the Higgs-$R^2$ model, the geometry is hyperbolic, so it is natural to think that this mechanism can occur. 

On super-horizon scales ($k \ll aH$), the effective isocurvature mass is written as 

\beq 
    m_\text{eff}^2 = N^I N^J\,\nabla_I V_J + H^2 \epsilon R_\text{fs} + 3(H\eta_\perp)^2
\enq

In the weak-coupling regime ($\xi_h \to 0$), the turning rate vanishes, thus we can approximate the effective mass as

\beq 
    m_\text{eff}^2 \simeq N^I N^J\,\nabla_I V_J + H^2 \epsilon R_\text{fs} = m_s^2+ \mu^2\,,
    \label{effective mass weak coupling}
\enq

where $m_s^2 = N^I N^J\,\nabla_I V_J$ is the conventional mass term and 
$\mu^2 = H^2 \epsilon R_\text{fs}$ is the geometrical contribution. 
At this point, a natural question arises: does geometrical destabilization 
play a significant role in the generation of isocurvature perturbations? 
In other words, does the field-space curvature term $R_\text{fs}$ 
significantly affect the perturbation dynamics in the Higgs--$R^2$ model?

To address this question, we note that during slow-roll inflation 
both $H$ and $\epsilon$ vary only slowly, so that the geometrical 
contribution $\mu^2$ can be treated as approximately constant 
during most of inflation. Under this approximation, the evolution 
equation for $Q_s$ in Eq.~(\ref{Qs ec.}) can be solved analytically 
in terms of conformal time $\tau$. In the weak-coupling limit ($\eta_\perp \to 0$), the simplified version of the action \eqref{second order action} can be written in terms of Mukhanov-Sasaki variables defined as $u = z\mathcal{R}_k$, $v = a Q_s$, then the action takes the form

\beq 
    S^{(2)} = \dfrac{1}{2}\int \dif x^3 \dif \tau \left[\left(u'^2 -(\nabla u)^2 + \dfrac{z''}{z}u^2\right) + \left(  v'^2 -(\nabla v)^2 + \dfrac{a''}{a}v^2 - m_\text{eff}^2\,a^2 v^2  \right)\right] \,,
\enq 

where $m_\text{eff}^2$ is given by \eqref{effective mass weak coupling}. For the isocurvature modes $Q_s$, the field equation in terms of $v$ and in the Fourier space reads as

\beq
    v''(\tau) + \omega_k^2(\tau)\,v(\tau) = 0\,.
\enq

Assumming quasi-De Sitter expansion, the time-dependent frequency $\omega_k^2$ is given by 

\beq
    \omega_k^2 = k^2 + \dfrac{1}{\tau^2}(\beta - 2 -\alpha)\,,
\enq

where $\alpha\,,\beta$ are dimensionless parameters defined as

\beq 
    \alpha = -\epsilon\,R_\text{fs}\,,\quad \beta = \dfrac{m_s^2}{H^2} 
\enq

The parameters relate the conventional mass term $m_s^2 = V_{NN}$ and the mass induced by the geometrical contribution of the field space, respectively. In the Higgs-$R^2$ model, $\alpha > 0$ since $R_\text{fs} <0$. Since the background evolution proceeds in the slow-roll regime,
both $H$ and $m_s^2$ vary slowly with time, so the parameter
$\beta = m_s^2/H^2$ can be treated as approximately constant
during horizon crossing. Thus, the resulting equation has the standard form of a Bessel differential equation, which allows us to express the general solution in terms of Hankel functions of the first kind, $H_\nu^{(1)}(-k\tau)$. The order of the Hankel function, $\nu$, is determined by

\beq
    \nu^2 = \frac{9}{4} + \alpha - \beta \,.
\enq

On super-horizon scales ($-k\tau \ll 1$), the amplitude of the isocurvature perturbations is governed by the asymptotic limit of the Hankel functions, yielding the standard scaling for the isocurvature power spectrum:

\beq
    \mathcal{P}_{\mathcal{S}} \propto k^{3-2\nu} \,.
\enq

 Given the hyperbolic geometry of our model ($R_\text{fs} = -1/3$), we find $\alpha = \epsilon / 3$. During the observable window of inflation where CMB scales exit the horizon ($N_e \sim 50-60$), the first slow-roll parameter is highly suppressed, typically $\epsilon \sim \mathcal{O}(10^{-4})$ (see Fig~\ref{Fig2}.). Consequently, the geometrical contribution to the effective mass is miniscule ($\alpha \ll 1$).

We can now explicitly compare this exact result against a hypothetical flat-space baseline ($R_\text{fs} = 0$). In the flat-space scenario, the geometrical term strictly vanishes ($\alpha = 0$), and the index reduces to $\nu_\text{flat}^2 = 9/4 - \beta$. Since the difference between the full hyperbolic geometry and the flat-space approximation is deeply bounded by $\epsilon/3$, the spectral index and the super-horizon amplitude of $\mathcal{P}_{\mathcal{S}}$ are virtually indistinguishable in both cases. 
This analytical treatment shows that the geometrical contribution is subdominant. The survival of the isocurvature modes in the weak-coupling regime is governed entirely by the extreme flatness of the potential landscape along the transverse direction (encoded in $\beta$), rather than by the field-space geometry. The geometrical contribution becomes relevant only toward the end of inflation, when the slow-roll approximation breaks down ($\epsilon \rightarrow 1$) and the field velocity increases.

\section{Impact on CMB observables}
\label{sec5}

In this section, we compute the angular power spectra for temperature and polarization anisotropies to illustrate the observational imprints of the multi-field dynamics discussed above. The presence of transient isocurvature modes and the non-trivial turning rate during inflation lead to distinctive deviations from the standard, featureless power-law spectrum predicted by single-field slow-roll inflation. We process the primordial power spectra $\mathcal{P}_{\mathcal{R}}(k)$ obtained in the previous section using the Boltzmann code \texttt{CLASS}~\cite{class}, assuming a standard $\Lambda$CDM background cosmology consistent with the Planck 2018 best-fit parameters~\cite{planck2018x}. The power spectra \eqref{powerspectrum} can be used to define the standard inflationary observables relevant to the CMB scales, which allows us to explore the large and intermediate scales relevant for low-$\ell$ suppression. The low-$\ell$ anomaly is well-known, albeit statistically mild, deficit of power observed in the CMB temperature anisotropies at large angular scales ($\ell \lesssim 40$) compared to the predictions of the base $\Lambda$CDM model. While often attributed to cosmic variance, transient dynamical features in multi-field inflation~\cite{Contaldi2003}---such as sharp turns in the field-space trajectory or brief departures from slow-roll---have been extensively shown to provide a compelling physical mechanism to naturally suppress the primordial power spectrum at these specific scales~\cite{Braglia2022a, Braglia2022b, Petretti_2025}.

\begin{table}[t!]
    \centering
    \begin{tabular}{c c c c c c c}
        \hline\hline
        $\xi_h$
        & $N_\text{hc}$
        & $P_\mathcal{R}(k_*)$
        & $n_s$
        & $r$
        & $\beta_\text{iso}$
        & $\cos\Delta$
        \\
        \hline
        $10^{-9}$
        & $50$
        & $2.40\times10^{-9}$
        & $0.961$
        & $4.26\times10^{-3}$
        & $1.01\times10^{-2}$
        & $0$
        \\
        $0.1$
        & $50$
        & $2.08\times10^{-9}$
        & $0.969$
        & $4.93\times10^{-3}$
        & $0$
        & $-0.78$
        \\
        \hline\hline
    \end{tabular}
    \caption{Inflationary and CMB-related observables for Higgs--$R^2$ inflation evaluated at the pivot scale $k_* = 0.05\,\mathrm{Mpc}^{-1}$. The parameters $\lambda = 10^{-10}$ and $\xi_s = 4\times10^8$ are kept fixed, while two representative values of the Higgs non-minimal coupling $\xi_h$ are shown.}
    \label{table1}
\end{table}

\subsection{Implications for angular CMB power spectra}
\label{subsec:cmb_spectra}

\begin{figure}[t]
    \centering
    \includegraphics[width=\textwidth]{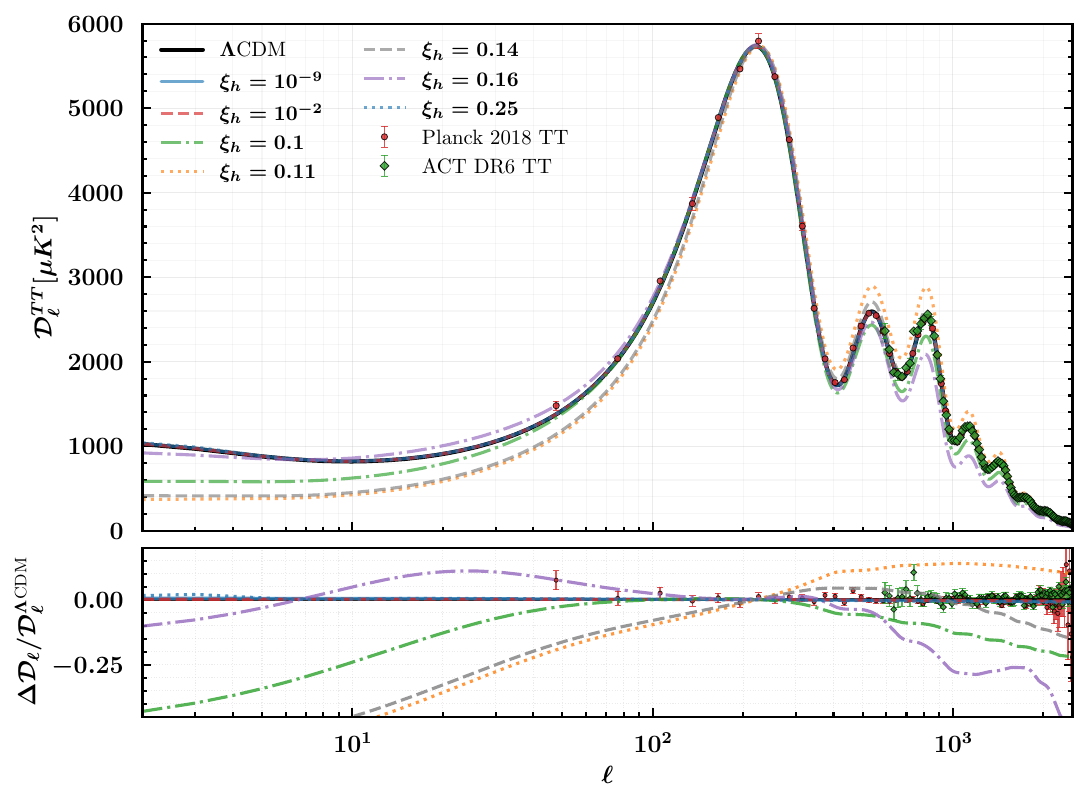}
   \caption{Top panel: Temperature angular power spectra $\mathcal{D}_\ell^{TT}$ for different values of the non-minimal coupling $\xi_h$, compared against Planck 2018 and ACT DR6 data. Bottom panel: Fractional differences relative to the baseline $\Lambda$CDM model.}
\label{fig:cmb_spectra}
    \label{fig7}
\end{figure}

To rigorously assess the phenomenological viability of the multi-field effects, we perform a fine-grained exploration of the parameter space around $\xi_h \sim \mathcal{O}(0.1)$, compared with the $\Lambda$CDM baseline, the weak-coupling regime case $\xi_h \ll 1$, and current CMB measurements from Planck~\cite{planck2018x} and ACT DR6~\cite{louis2025} shown for reference. As depicted in Fig.~\ref{fig7}, the non-minimal coupling $\xi_h$ effectively controls the multi-field dynamics, allowing us to identify three distinct regimes:

\begin{itemize}
    \item \textbf{Weak-coupling regime ($\xi_h \le 10^{-2}$):} The adiabatic spectrum is largely unaffected by the multi-field turn, remaining consistent with standard single-field predictions and maintaining an excellent fit to the CMB data. However, as discussed in Sec.~\ref{iso remain}, residual isocurvature modes survive until the end of inflation.
    
    \item \textbf{Intermediate regime ($0.1 \lesssim \xi_h \lesssim 0.5$):}
    This window is characterized by a strong transient turning of the trajectory.
    While this generates intriguing features at large scales ($\ell \lesssim 40$),
    the sustained power transfer mechanism induces a suppression of the adiabatic
    power at small scales. As seen in the residuals of Fig.~\ref{fig7}, values around
    $\xi_h \sim 0.1$ exhibit a significant deficit with respect to ACT DR6
    data at $\ell \gtrsim 1000$, in clear tension with the measurements.
    However, this discrepancy progressively weakens as $\xi_h$ increases,
    with nearby values showing a much better agreement with observations.

    \item \textbf{Effective single-field regime ($\xi_h \gtrsim 1$):} For large couplings, the multi-field turn is highly suppressed, and the isocurvature perturbations decay efficiently. The model recovers the standard, featureless adiabatic spectrum, bringing the high-$\ell$ tail back into perfect agreement with observations.
\end{itemize}

\begin{figure}[t]
    \centering
    \includegraphics[width=0.48\textwidth]{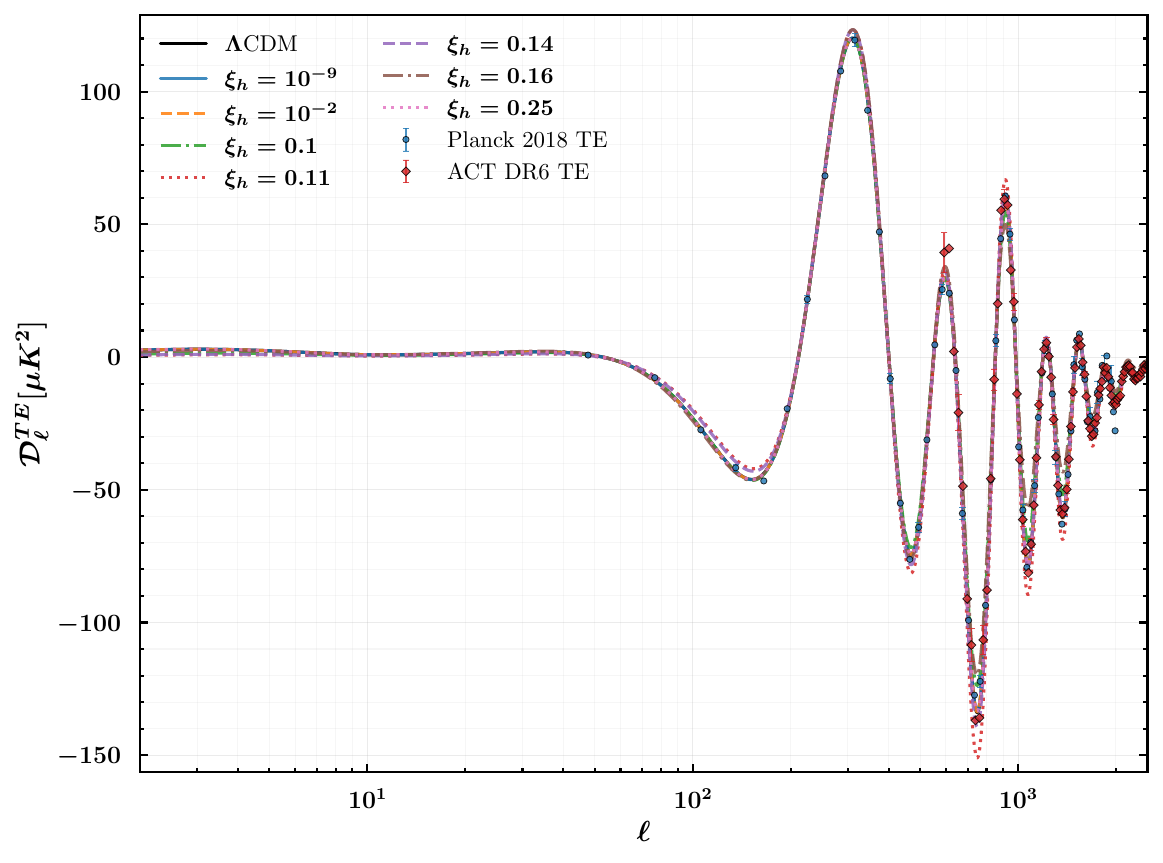}\hfill
    \includegraphics[width=0.48\textwidth]{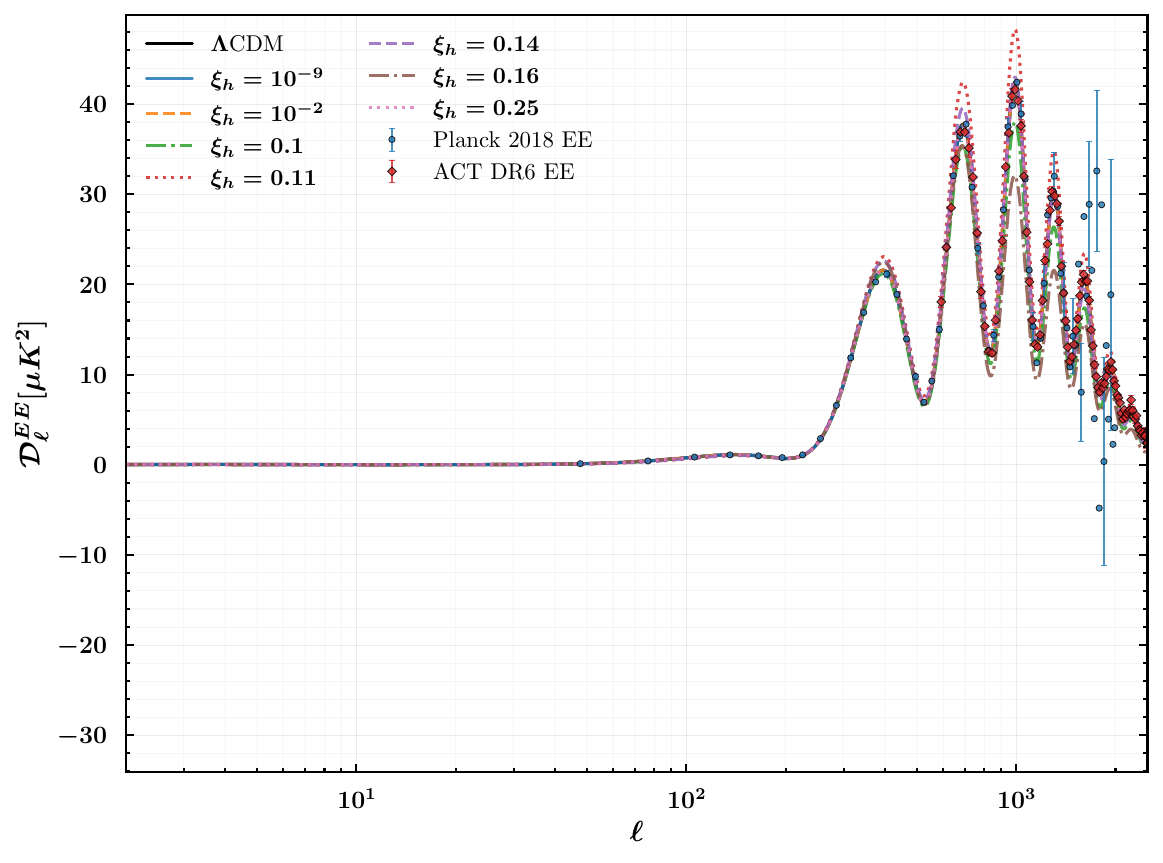}
  \caption{Temperature-polarization cross-correlation (TE, left) and E-mode polarization (EE, right) angular power spectra for selected values of the non-minimal coupling $\xi_h$. The theoretical predictions are compared against Planck 2018 and ACT DR6 data.}
    \label{fig8}
\end{figure}

The physical origin of these regimes can be understood in terms of the scale-dependent transfer between adiabatic and isocurvature modes. At large angular scales ($\ell \lesssim 40$), the coupling between adiabatic and isocurvature perturbations induced by the turning trajectory leads to a suppression of power in the Sachs--Wolfe plateau~\cite{Jain2009}. This effect overlaps with the multipole range probed by current temperature measurements and may partially alleviate the observed low-$\ell$ power deficit. However, it should
be interpreted as a qualitative imprint of the inflationary dynamics rather than a quantitative resolution of the anomaly. At small scales ($\ell \gtrsim 1000$), the same transfer mechanism induces a coherent suppression of power across the TT, TE, and EE spectra (see Fig.~\ref{fig8}), reflecting a scale-dependent
redistribution of primordial curvature perturbations. This behavior highlights the tension between generating large-scale features via multi-field dynamics and maintaining consistency with high-precision
CMB observations.

Therefore, we conclude that within this minimal setup, the intermediate
coupling regime is strongly constrained by current observations, and
primarily serves to illustrate the stringent limits that small-scale CMB
data impose on non-canonical multi-field inflationary models.

\section{Estimate of primordial non-Gaussianity}
\label{sec6}

Transient turns in multi-field inflation can generate non-Gaussian
signatures through the transfer of isocurvature fluctuations into
the adiabatic mode. The amplitude and shape of the bispectrum are
typically controlled by the turning rate $\eta_\perp$ and by the
duration of the conversion process.
A preliminary numerical exploration of the Higgs--$R^2$ model
suggests that the resulting non-Gaussianities remain moderate.
For equilateral configurations, the non-linearity parameter
approaches a small value at the end of inflation,
$f_{\rm NL}^{\rm equil} = 0.0159$, consistent with the
single-field expectation. During the transient turn,
temporary enhancements can occur with amplitudes
$f_{\rm NL}^{\rm equil} \sim \mathcal{O}(1$--$10)$, but these
features decay rapidly as the system returns to the universal
attractor. In contrast, the local configuration can reach amplitudes
$f_{\rm NL}^{\rm local} \sim \mathcal{O}(1$--$10)$ at the end of inflation, depending on
the initial conditions, reflecting the super-horizon conversion
of isocurvature modes into curvature perturbations.

A detailed analysis of the bispectrum in the Higgs--$R^2$ model
is currently in progress and will be reported elsewhere.

\section{Discussion and conclusions}
\label{sec7}

In this work, we have presented a comprehensive analysis of primordial perturbations in multi-field Higgs--$R^2$ inflation. By going beyond the effective single-field approximation and performing a full numerical integration of the background and perturbation equations, we uncovered a rich phenomenology controlled by the non-minimal coupling $\xi_h$. Our results demonstrate that the field-space geometry and the turning rate of the inflationary trajectory play a decisive role in shaping the primordial perturbations, leading to signatures that are absent in the standard single-field limit.
We identified two qualitatively distinct dynamical regimes depending on the magnitude of the Higgs non-minimal coupling. For weak coupling ($\xi_h \ll 1$), the turning rate $\eta_{\perp}$ remains negligible and the adiabatic power spectrum is essentially featureless. However, contrary to the standard heavy-field intuition, the isocurvature mode $Q_s$ remains light ($m_\text{iso}^2 \lesssim H^2$) and does not fully decay, leaving a residual isocurvature fraction $\beta_\text{iso} \sim \mathcal{O}(10^{-2})$ at the end of inflation. This implies that in the weak-coupling limit Higgs--$R^2$ inflation cannot be regarded as effectively single-field, but rather constitutes a genuinely multi-field scenario with surviving isocurvature perturbations that could be constrained by future high-precision observations.

A second and phenomenologically more distinctive regime emerges for $\xi_h \sim 0.1$. In this case, the inflationary trajectory undergoes a transient turn as it evolves from the ridge at $h=0$ toward the valley, resulting in a significant turning rate and a strong coupling between adiabatic and isocurvature modes. We showed that this interaction generates a localized suppression in the primordial curvature spectrum at large scales, accompanied by oscillatory features. Importantly, in this regime the isocurvature modes eventually become heavy and decay exponentially, yielding a purely adiabatic spectrum at the end of inflation.

A more detailed exploration of the parameter space around this intermediate regime reveals that the non-minimal coupling $\xi_h$ effectively controls the strength of the multifield effects. While values $\xi_h \sim 0.1$ lead to a significant suppression of power at small angular scales ($\ell \gtrsim 1000$), in tension with ACT DR6 data, this discrepancy progressively weakens for slightly larger values of $\xi_h$, for which the angular power spectra approach the $\Lambda$CDM prediction. This behavior indicates the existence of a nearby parameter region where large-scale features can be generated while maintaining consistency with small-scale observations, highlighting the strong constraining power of high-$\ell$ CMB measurements.

On the theoretical side, we have also clarified the role of the field-space geometry in the dynamics of isocurvature perturbations. Although the hyperbolic geometry induces a negative contribution to the effective isocurvature mass through the field-space curvature, we showed analytically that this geometrical term remains suppressed during the observable window of inflation, being proportional to the slow-roll parameter $\epsilon$. As a result, the evolution of isocurvature modes is predominantly governed by the potential curvature rather than by geometrical destabilization effects, which only become relevant near the end of inflation.

Given the large cosmic variance affecting multipoles $\ell \lesssim 40$, establishing a statistical preference for this scenario over the standard $\Lambda$CDM model would require a dedicated likelihood analysis in the low-$\ell$ regime. However, the constraints from smaller angular scales are significantly more restrictive. Even without a full parameter estimation, the systematic deficit of power at $\ell \gtrsim 1000$ for $\xi_h \sim 0.1$ exceeds the experimental uncertainties of the ACT DR6 measurements~\cite{louis2025}, placing this benchmark scenario under significant tension with current data. A definitive assessment of its viability would require a full likelihood analysis, which we leave for future work.

Finally, since the suppression of power in the acoustic peaks region in our model originates from transient multi-field effects associated with a non-trivial turn in field space, it is natural to ask whether the same dynamics could leave imprints in higher-order correlation functions. In particular, the interplay between adiabatic--isocurvature mixing and the turning rate suggests that this setup may lead to distinctive non-Gaussian signatures. Preliminary results indicate that while equilateral configurations remain small at the end of inflation, local-type non-Gaussianity can reach amplitudes of order $f_{\mathrm{NL}} \sim \mathcal{O}(1\text{--}10)$ and depends sensitively on the initial conditions. A detailed analysis of primordial bispectra will be presented in future work.

\section{Data Availability}

The numerical codes and data generated to solve the multi-field background equations and produce the figures in this study are publicly available in the GitHub repository at Ref.~\cite{pineda_data_2026}.

\section{Acknowledgments}

This work was partially supported by SECIHTI under grant CBF2023--2024--1937. FP acknowledges financial support from SECIHTI--México under grant No.~803062.

\appendix
\section{Single-field inflation along the valleys
\label{appendix1}}

The valley approximation of the potential \eqref{potential Einstein} in the Einstein frame is relevant for analyzing the evolution of the fields $(\phi, h)$. This analysis has been discussed in detail in previous works on this model \cite{Wang2017,GUNDHI2020,He2018}; therefore, here we restrict ourselves to outlining the main ideas. In the limit $h\gg v_\text{ew}$, the valleys of the potential \eqref{potential Einstein} are defined by the condition

\beq
\dfrac{\partial V}{\partial h} = 0\,,\quad h_0^2(\phi) = \dfrac{\xi_h}{\xi_h^2 + 4\lambda \xi_s}(e^{\alpha \phi} - 1 )
\label{valley}
\enq

This condition represents a stationary point in the $(h, \dot{h})$ plane; as the scalaron $\phi$ rolls down to the minimum of the potential \eqref{potential Einstein}, the valleys act as a universal attractor. Since the potential is symmetric under $h\to - h$, Eq. \eqref{valley} identifies the two valleys of $V(\phi, h)$. Along one of the valleys, the potential is given by the following effective potential as a function of $\phi$:

\beq
    V_\text{eff}(\phi\,,h_0(\phi)) = \dfrac{ \lambda}{4(\xi_h^2 + 4\lambda\xi_s)}(1 - e^{-\alpha\phi})^2\,.
    \label{effective potential}
\enq

This potential has exactly the same form as the potential in $R^2$ inflation or the Higgs inflation model \cite{BEZRUKOV2008}. For the standard parameter space discussed above, the model effectively reduces to a single scalar field model. Given that the Higgs field along the valleys is determined by \eqref{valley}, the action \eqref{Higgs-R2 action} along these trajectories takes the form

\beq
    S_E[g_\mu\,,\phi\,,h] = \int \dif^4 x\,\sqrt{-g}\,\left[\dfrac{1}{2}R - \dfrac{1}{2}g^{\mu\nu}\partial_\mu\phi\partial_\nu\phi \left(1 + \gamma^2(1 - e^{-\alpha\phi})^{-1}\right) -V_\text{eff}(\phi)\right]\,,
\enq

where $\tilde{\xi}^2 = \frac{\xi_h}{6(\xi_h^2 + 4\lambda\xi_s)}$. We observe that if $\lambda = 0.13$ and $\xi_h \sim 10^4$, the constant is of the order $\tilde{\xi} \sim 10^{-5}$; thus, the second factor in the kinetic term becomes subdominant. At the inflationary scale $\phi \gg 1$, the model effectively reduces to canonical single-field inflation with the effective potential given by \eqref{effective potential}. However, for a parameter space such that $\xi_h \sim \lambda\xi_s$, the constant becomes $\tilde{\xi}\sim \mathcal{O}(1)$. This case corresponds to what is known as the quasi-single field inflation regime \cite{chen2010prd,Chen_2010}. Along the valleys, both the unit vectors and the turning rate adopt a simplified form:

\beq
    T^I = \dfrac{(1\,, \tilde{\xi} e^{\alpha\phi/2})}{\sqrt{1 + \tilde{\xi}^2}}\,,\quad N^I = \dfrac{(-\tilde{\xi}\,,  e^{\alpha\phi/2})}{\sqrt{1 + \tilde{\xi}^2}}\,, \quad \eta_\perp = \dfrac{\tilde{\xi}}{\sqrt{1 + \tilde{\xi}^2}}
    \label{valley approach}
\enq

If $\tilde{\xi} \ll 1$, the vector $T^I$ aligns with the $\phi$ direction, while $N^I$ points in the $h$ direction. The turning rate $\eta_\perp$ is negligible within the valleys.

\end{document}